\documentclass[11pt]{article} 
\usepackage{amssymb}
\usepackage{amsmath}
\usepackage{setspace}
\usepackage{amsthm}
\usepackage{epsfig,color,colordvi,wasysym}
\usepackage{graphicx}

\usepackage[top=.8in,bottom=1in,left=1in,right=1in]{geometry}

\title{Fair Evaluation of Global Network Aligners}
\author{Joseph Crawford,\thanks{Department of Computer Science and Engineering, University of Notre Dame} \thanks{Interdisciplinary Center for Network Science and Applications
(iCeNSA), University of Notre Dame} \thanks{ECK Institute for Global Health, University of Notre Dame} \emph{}
Yihan Sun,\footnotemark[1] \thanks{Department of Computer Science and Technology, Tsinghua University} \emph{} and
Tijana Milenkovi\'{c}\footnotemark[1] \footnotemark[2] \footnotemark[3] \thanks{Corresponding Author (E-mail: tmilenko@nd.edu)}
}

\begin{document}
\maketitle

\begin{abstract} % abstract
Analogous to genomic sequence alignment, biological network alignment
identifies conserved regions between networks of different
species. Then, function can be transferred from well- to
poorly-annotated species between aligned network regions. Network
alignment encompasses two algorithmic components: node cost function
(NCF), which measures similarities between nodes in different
networks, and alignment strategy (AS), which uses these similarities
to rapidly identify high-scoring alignments. Different methods use
both different NCFs and different ASs. Thus, it is unclear whether the
superiority of a method comes from its NCF, its AS, or both. We
already showed on state-of-the-art methods, MI-GRAAL and IsoRankN,
that combining NCF of one method and AS of another method \emph{can}
lead to a new superior method. Here, we evaluate MI-GRAAL against a
newer approach, GHOST, by mixing-and-matching the methods' NCFs and
ASs to potentially further improve alignment quality. While doing so,
we approach several important questions that have not been asked
systematically thus far. First, we ask how much of the node similarity
information within NCF should come from protein sequence data compared
to network topology data. Existing methods determine this parameter
more-less arbitrarily, which could significantly affect the resulting
alignment(s). Second, when topological information is used in NCF, we
ask how large the size of the neighborhoods of the compared nodes
should be. Existing methods assume that the larger the neighborhood
size, the better.

We find that MI-GRAAL's NCF is superior to GHOST's NCF, while the
performance of the methods' ASs is data-dependent. Thus, for data on
which GHOST's AS is superior to MI-GRAAL's AS, the combination of
MI-GRAAL's NCF and GHOST's AS represents a new superior method. Also,
we find that which amount of sequence information is used within NCF
does not affect alignment quality, while the inclusion of topological
information is crucial for producing good alignments. Finally, we
find that larger neighborhood sizes are preferred, but often, it is
the second largest size that is superior. Using this size instead of
the largest one would decrease computational complexity.

Taken together, our results lead to several general recommendations
for a fair evaluation of network alignment methods.

\end{abstract}

%%%%%%%%%%%%%%%%%%%%%%%%%%%%%%%%%%%%%%%%%%%%%%
%% %%
%% The keywords begin here %%
%% %%
%% Put each keyword in separate \kwd{}. %%
%% %%
%%%%%%%%%%%%%%%%%%%%%%%%%%%%%%%%%%%%%%%%%%%%%%

% MSC classifications codes, if any
%\begin{keyword}[class=AMS]
%\kwd[Primary ]{}
%\kwd{}
%\kwd[; secondary ]{}
%\end{keyword}

\section{Background} \label{intro}

\subsection{Motivation and related work}

Analogous to sequence alignment, which finds regions of similarity
that are a likely consequence of functional or evolutionary
relationships between the sequences, network (or graph) alignment
finds regions of topological and functional similarity between
networks of different species \cite{Sharan2006}. Then, functional
(e.g., aging-related
\cite{MilenkovicACMBCB2013,Faisal2014a,Faisal2014}) knowledge can be
transferred between species across conserved (aligned) network
regions. Thus, just as sequence alignment, network alignment can be
used for establishing from biological network data orthologous
relationships between different proteins or phylogenetic relationships
between different species \cite{GRAAL,HGRAAL,MIGRAAL}. Also, it can be
used to semantically match entities in different ontologies
\cite{li2009rimom}.

Network alignment can be performed locally and globally. Local network
alignment (LNA) aims to optimize similarity between local regions of
different networks
\cite{PathBlast,Sharan2005,Flannick2006,Mawish,Berg04,Liang2006a,Berg2006,Mina20014,AlignNemo}.
As such, LNA often leads to many-to-many node mapping between
different networks. However, LNA is generally unable to find large
conserved subgraphs. Thus, methods for global network alignment (GNA)
have been proposed, which aim to optimize global similarity between
different networks and can thus find large conserved subgraphs
\cite{Singh2007,Flannick2008,Singh2008,GraphM,IsoRankN,GRAAL,HGRAAL,MIGRAAL,GHOST,NETAL,Narayanan2011,Guo2009,NATALIE,NATALIE2,IsoRankN,MilenkovicACMBCB2013,Faisal2014a,MAGNA,NewSurvey2014}.
Unlike LNA, GNA typically results in one-to-one node mapping between
different networks. In this study, we focus on GNA due to its recent
popularity, but all concepts and ideas can be applied to LNA as well.

More formally, we define GNA as a one-to-one mapping between nodes of
two networks that aligns the networks well with respect to a desired
topological or functional criterion. GNA is a computationally hard
problem to solve due to the underlying subgraph isomorphism problem
\cite{West96}. This is an NP-complete problem that asks whether a
network exists as an exact subgraph of a larger network. GNA is a more
general problem which aims to fit well two networks when one network
is not necessarily an exact subgraph of another network. Since GNA is
computationally hard, heuristic methods need to be sought. Existing
GNA heuristic algorithms typically achieve an alignment via two
algorithmic components: node cost function (NCF) and alignment
strategy (AS). NCF captures pairwise costs (or equivalently,
similarities) of aligning nodes in different networks, and AS uses
these costs to identify a good-quality alignment out of all possible
alignments with respect to some topological or biological alignment
quality measure.

Different existing GNA methods generally use both different NCF and
AS, so it is unclear whether the superiority of a method comes from
its NCF, AS, or both. For this reason, in our recent study
\cite{MilenkovicACMBCB2013,Faisal2014a}, we combined 
NCFs and ASs of MI-GRAAL \cite{MIGRAAL} and IsoRankN \cite{IsoRankN},
two state-of-the-art methods at the time, as a proof of concept that
it is important to fairly evaluate the contribution of each component
to alignment quality. In the process, we showed that NCF of MI-GRAAL
is superior to that of IsoRankN, and importantly, we proposed the
combination of MI-GRAAL's NCF and IsoRankN's AS as a new superior
method for multiple GNA, i.e., for GNA of more than two networks at a
time \cite{MilenkovicACMBCB2013,Faisal2014a}.

In the meanwhile, a new state-of-the-art method has appeared, called
GHOST \cite{GHOST} (along with some other methods
\cite{NewSurvey2014}). Thus, in this study, we fairly evaluate
MI-GRAAL against GHOST by mixing and matching their NCF and AS. At the
same time, we ask several additional important questions regarding the
choice of appropriate GNA parameters, which have surprisingly been
neglected thus far.

\subsection{Our approach and contributions} \label{approach} 

MI-GRAAL \cite{MIGRAAL} and GHOST \cite{GHOST} are two
state-of-the-art global network aligners that injectively map nodes
between two networks in a way that preserves topologically or
functionally conserved network regions. The two methods are
conceptually similar, in the sense that their NCFs assume two nodes
from different networks to be similar if their topological
neighborhoods are similar. However, the mathematical and
implementation details of the two NCFs are different. The same holds
for the two methods' ASs. To evaluate the contribution to the
alignment quality of each of the two NCFs and two ASs, we mix and
match these, resulting in a total of four different combinations. We
then use each combination to produce alignments for synthetic networks
with known ground truth node mapping as well as for real-world
networks without known ground truth node mapping, and we evaluate the
quality of each alignment with respect to five topological and two
biological alignment quality measures.

In general, we find that MI-GRAAL's NCF is superior to GHOST's NSF,
while the superiority of the methods' ASs is data-dependent. Hence,
for those network data on which GHOST's AS is superior to MI-GRAAL's
AS, we propose the combination of MI-GRAAL's NCF and GHOST's AS as a
new superior network aligner.

While fairly evaluating MI-GRAAL's and GHOST's NCFs and ASs, we
approach two additional important research questions that, to our
knowledge, have not been asked systematically in the context of
network alignment thus far: 1) how much of the node similarity
information within the NCF should come from protein sequence data
compared to network topology data, and 2) how large the size of the
neighborhoods of the compared nodes from different networks should be
when generating topological similarity information within the
NCF. Current GNA methods generally use a seemingly arbitrary amount of
sequence information in their NCF, and also, they assume that the
larger the size of a node's neighborhood, the better the alignment
quality. Thus, in this study, we evaluate whether these
``state-of-the-art'' choices are actually appropriate. 

In general, we find that which amount of sequence information is used
within NCF does not drastically affect neither topological or
biological alignment quality, while the effect of topological
information is drastic. Namely, using no topological information
within NCF results in poor topological and sometimes even biological
alignment quality. Hence, topology takes precedence over sequence when
it comes to improving alignment quality. Also, we find that using
larger network neighborhood sizes within NCF in most cases leads to
better alignment quality than using smaller neighborhood sizes.
However, it is not always the case that the largest neighborhood size
is the best; in many cases, the second largest size is the best.
Therefore, using this size instead of the largest one would
drastically decrease computational complexity of the given method
without decreasing its accuracy.

\section{Methods}

\subsection{Data sets} \label{data}

We use two popular benchmark sets of networks in this study: 1)
synthetic networks with known ground truth node mapping and 2)
real-world protein-protein interaction (PPI) networks without known
ground truth node mapping
\cite{MIGRAAL,GHOST,MAGNA,MilenkovicACMBCB2013,Faisal2014a}.

The synthetic network data with known node mapping consists of a
high-confidence yeast PPI network, which has 1,004 proteins and 8,323
PPIs \cite{Collins07,GRAAL, HGRAAL, MIGRAAL, GHOST, MAGNA}, and five
additional networks that add noise to the yeast network. Noise is the
addition to the yeast network of low-confidence edges from the same
data set \cite{Collins07}, and each of the five additional noisy
networks adds $x\%$ noise to the original network, where $x$ varies
from 5\% to 25\% in increments of 5\%. In this network set, we align
the original yeast network to each of the synthetic networks with
$x\%$ noise, resulting in the total of five network pairs to be
aligned.

The real-world PPI network data without known node mapping consists of
PPI networks of the following four species: \emph{S. cerevisiae}
(yeast/Y), \emph{D. melanogaster} (fly/F), \emph{C. elegans} (worm/W),
and \emph{H. sapiens} (human/H). The yeast, fly, worm, and human
networks have 3,321 proteins and 8,021 PPIs, 7,111 proteins and 23,376
PPIs, 2,582 proteins and 4,322 PPIs, and 6,167 proteins and 15,940
PPIs, respectively \cite{BIOGRID}. In this network set, we align PPI
networks for each pair of species, resulting in the total of six
network pairs to be aligned.

We note that the synthetic network data is not truly synthetic, as
both the original yeast network and the noise in terms of the
lower-confidence PPIs come from an actual experimental study
\cite{Collins07}. We refer to this network set as synthetic simply
because we know the known ground truth node mapping, unlike for the
real-world PPI network set. Also, we note that the synthetic network
data encompasses ``co-complex'' PPIs obtained by affinity purification
followed by mass-spectrometry (AP/MS), among other PPI types, while
the real-world PPI network data consists of ``binary'' yeast
two-hybrid (Y2H) PPIs. Another difference between the two network sets
is for the synthetic data the smaller (original yeast) network is an
exact subgraph of the larger (noisy) network, whereas this is not the
case for networks of different species in the real-world data.

For evaluating the amount of sequence that should be used within NCF
when generating an alignment, we use protein sequence similarity data.
This data comes from BLAST bit-values from the NCBI database
\cite{Altschul90}.

When evaluating the biological alignment quality with respect to
functional enrichment of the aligned nodes, we use Gene Ontology (GO)
annotation data from our recent study
\cite{MilenkovicACMBCB2013,Faisal2014a}, to allow for fair and
consistent method evaluation.

\subsection{Existing network aligners and their NCFs and ASs} \label{aligners}

\noindent{\bf MI-GRAAL's NCF.} MI-GRAAL improves upon its
predecessors, GRAAL \cite{GRAAL} and H-GRAAL \cite{HGRAAL}, by using
the same NCF (see below) but by combining GRAAL's and H-GRAAL's ASs
into a new superior AS (see below).

\begin{figure}[h]
\centering
\includegraphics[height=2.7in]{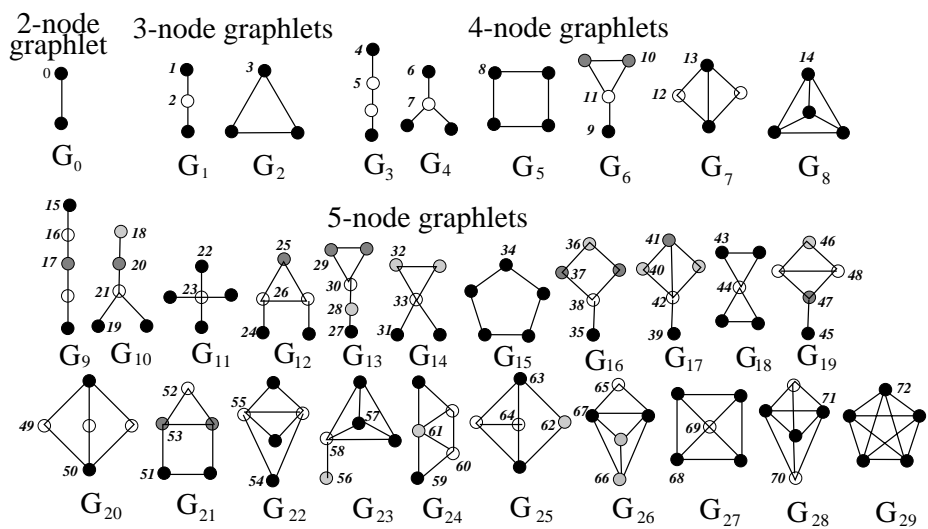}
\caption{Illustration of MI-GRAAL's NCF, which uses the thirty 2-, 3-,
4-, and 5-node graphlets $G_0, G_1, ..., G_{29}$ and their ``node
symmetry groups'', also called automorphism orbits, numbered 0, 1,
2, ..., 72. In a graphlet $G_i$, $i \in \{0,1,...,29\}$, nodes
belonging to the same orbit are of the same shade. For details, see
the original publication \cite{Przulj06ECCB}.}
\label{fig:graphlet}
\end{figure}

MI-GRAAL's NCF relies on the concept of small induced subgraphs called
$graphlets$ (Fig. ~\ref{fig:graphlet})
\cite{Milenkovic2008,GraphCrunch,MMGP_Roy_Soc_09,Milenkovic2011,Solava2012,Hulovatyy2014}.
All 2-5-node graphlets are considered. Because of the small-world
nature of real-world networks, using larger graphlets would
unnecessarily increase the computational complexity needed the count
the graphlets \cite{GRAAL,HGRAAL}. Based on the graphlets, the
\emph{node graphlet degree vector} (node-GDV) is computed for each
node in each network, which counts how many times the given node
touches each of the 2-5-node graphlets, i.e., each of their 73 node
symmetry groups (or \emph{automorphism orbits}; Fig.
~\ref{fig:graphlet}). As such, node-GDV captures up to a 4-deep
network neighborhood of the node of interest. By comparing node-GDVs
of two nodes to compute their \emph{node-GDV-similarity}, and by doing
so between each pair of nodes in different networks, one is able to
capture pairwise topological node similarities between the different
networks. MI-GRAAL also allows for integration of other node
similarity measures into
its NCF, such as protein sequence similarity.\\

\noindent{\bf MI-GRAAL's AS.} GRAAL's AS utilizes a seed-and-extend 
approach to greedily maximize the total NCF over all aligned nodes.
H-GRAAL, on the other hand, finds optimal alignments with respect to
the total NCF by using the Hungarian algorithm to solve the linear
assignment problem. MI-GRAAL's AS combines GRAAL's greedy
seed-and-extend approach with H-GRAAL's optimal AS into a superior
AS. For details on MI-GRAAL's AS, see the original publication
\cite{MIGRAAL}.\\

\begin{figure}[h]
\centering
\includegraphics[width=4in]{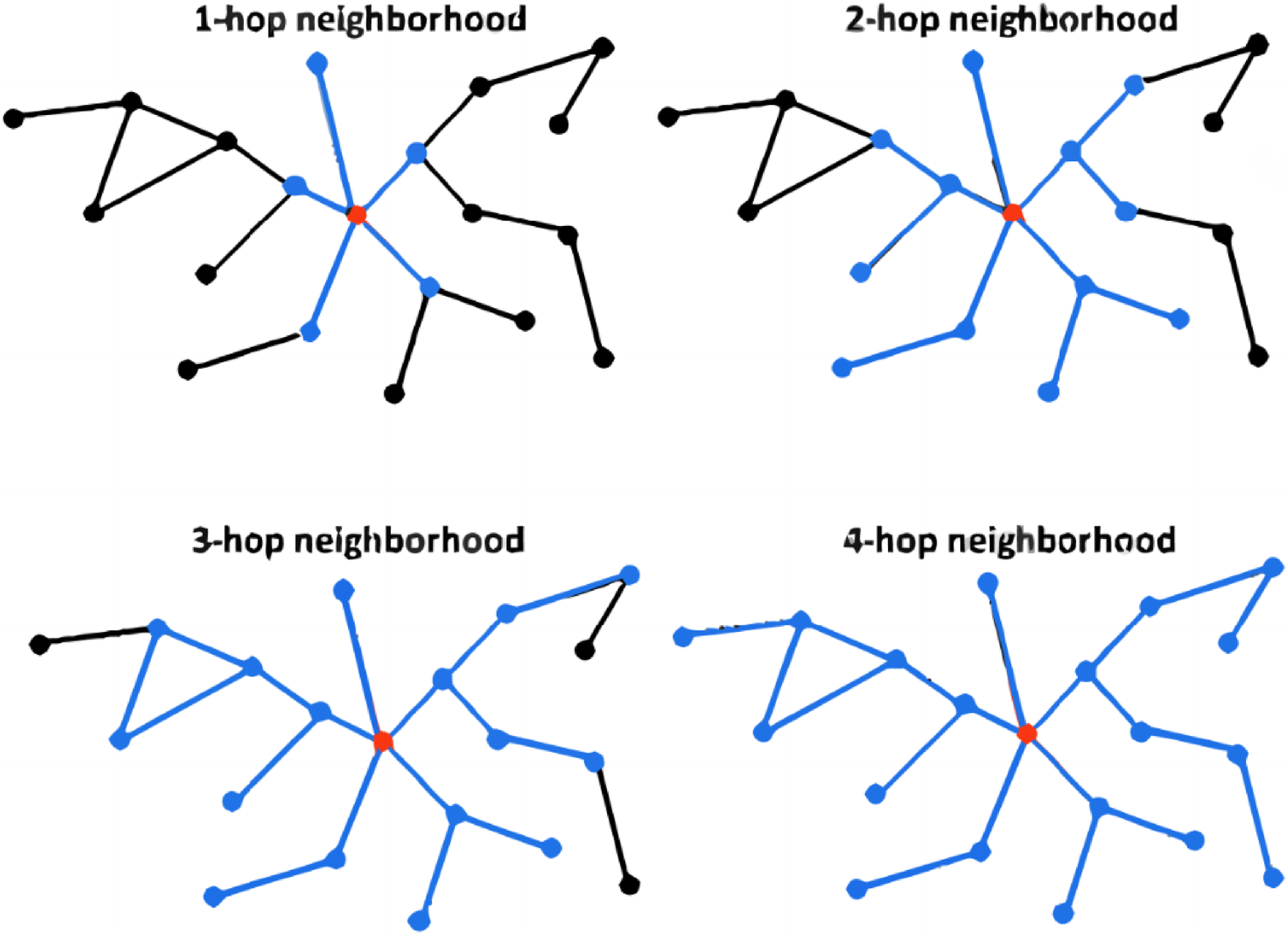}
\caption{Illustration of GHOST's NCF, which compares two nodes in
different networks with respect to similarity of each of their
$k$-hop neighborhoods, $k=1,2,3,4$. All blue edges and blue nodes
are within the given $k$-hop neighborhood of the red node.}
\label{fig:khop}
\end{figure}

\noindent{\bf GHOST's NCF.} GHOST's NCF takes into account a node's
$k$-hop neighborhood ($k=4$), which is the induced subgraph on all
nodes whose shortest path distance from the node in question is less
than or equal to $k$ (Fig. ~\ref{fig:khop}). Intuitively, GHOST's NCF
computes topological distance (or equivalently similarity) between two
nodes from different networks by comparing the nodes' ``spectral
signatures'', where the spectral signature of a node is based on
subgraph counts in the node's the $k$-hop neighborhood \cite{GHOST}.
For more details on GHOST's NCF, see the original publication
\cite{GHOST}. In our study, we consider $k=1, 2, 3, 4$, which allows
for a fair comparison of GHOST's NCF to MI-GRAAL's NCF when varying
the size of network neighborhood that is considered within the NCFs
(Section ~\ref{generate}).

\begin{figure}[h]
\centering
\includegraphics[width=2in]{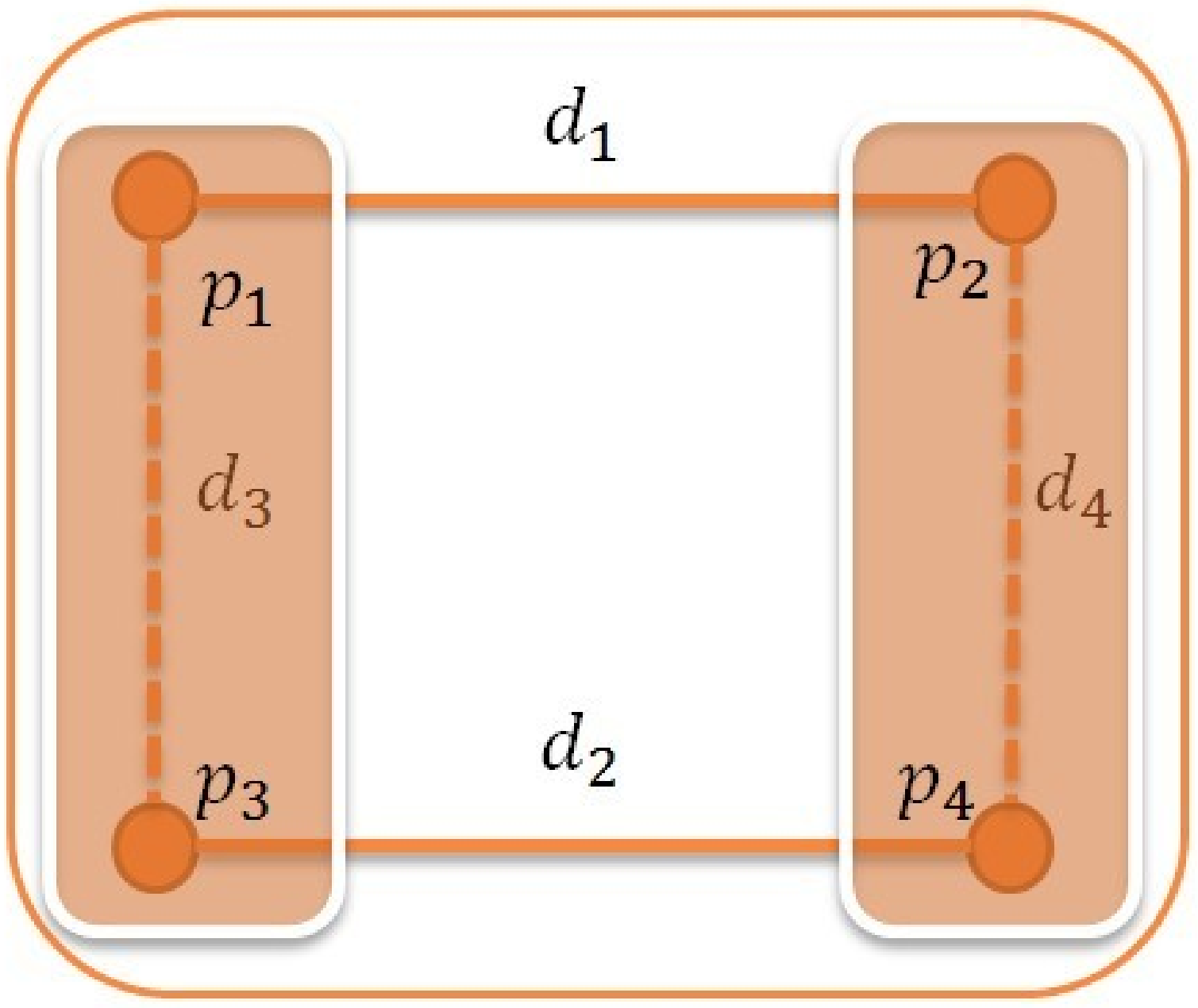}
\caption{Intuitive comparison of MI-GRAAL's and GHOST's ASs. Let us
assume that we are aligning two graphs $G_1(V_1,E_1)$ and
$G_2(V_2,E_2)$. Let $p_1,p_3 \in V_1$, let $p_2,p_4 \in V_2$, and
let the NCF distance (er equivalently, similarity) between the node
pairs be $d_1,d_2,d_3,d_4$, as illustrated. MI-GRAAL's alignment
strategy only considers the values $d_1$ and $d_2$ when creating an
alignment, while GHOST's AS considers the values $d_1$,$d_2$,$d_3$,
and $d_4$ when doing so.}
\label{fig:distance}
\end{figure}

\noindent{\bf GHOST's AS.} Unlike MI-GRAAL's AS that solves linear
assignment problem, GHOST's AS solves the quadratic assignment problem
(Fig.~\ref{fig:distance} illustrates this). For further details on
GHOST's AS, refer to the original publication \cite{GHOST}.

\subsection{Aligners resulting from combining existing NCFs and ASs, and their parameters} \label{generate}

\noindent{\textbf{Mixing and matching different NCFs and ASs.}} To
fairly evaluate the two algorithmic components of MI-GRAAL and GHOST,
we aim to first compare the two NCFs under the same AS, for each of
the two ASs. We then aim to compare the two ASs under the same NCF,
for each of the two NCFs. This results in a total of four aligners,
i.e., different combinations of the two methods' NCFs and ASs.
However, GHOST does not allow the user to import their own (e.g.,
MI-GRAAL's) NCF into its AS, so we are unable to study the combination
of MI-GRAAL's NCF and GHOST's AS. Thus, in total, we consider three
different aligners (Table ~\ref{tab:aligner}).

\begin{table}[h]
\centering
\begin{tabular}{| l | l | l |}
\hline
Aligner & Node Cost Function & Alignment Strategy \\
\hline
M-M & MI-GRAAL & MI-GRAAL \\
\hline
G-M & GHOST & MI-GRAAL \\
\hline
G-G & GHOST & GHOST \\
\hline
\end{tabular} 
\vspace{0.1cm}
\caption{The three aligners considered in this study. The first letter 
in the aligner represents NCF of the aligner, while the second letter
represents AS of the aligner.}
\label{tab:aligner}
\end{table}

\noindent{\textbf{Varying the amount of sequence versus topological information within NCF.}}
An additional goal of this paper is to determine the most appropriate
amount of sequence information versus topological information to be
included into NCF. Thus, for each aligner, we generate NCFs with
varying amounts of sequence and topology information, as $\alpha T +
(1-\alpha)S$, where $T$ represents topological similarity score (e.g.
node-GDV-similarity) and $S$ represents sequence similarity score. We
vary $\alpha$ from 0 to 1 in increments of 0.1.

\noindent{\textbf{Varying the size of network neighborhood within NCF.}}
Further, we aim to determine the most appropriate neighborhood size
that should be used within NCF when producing an alignment. Thus, for
each aligner (and for each value of $\alpha$), we also consider four
different neighborhood sizes, as described in Table~\ref{tab:top}. We
note that although we have tried to classify under the same
neighborhood size label (e.g. T1 in Table~\ref{tab:top}) graphlet
sizes considered within MI-GRAAL's NCF and $k$-hop values considered
within GHOST's NCF, we note that it is not necessarily the case that
the neighborhood of a node that is covered by graphlets of a given
size and the neighborhood of the same node that is covered by the
corresponding $k$-hop value match exactly. That is, for example,
2-3-node graphlets and 2-hop neighborhood (both corresponding to T2 in
Table~\ref{tab:top}) do not necessarily cover exactly the same amount
of network topology. Yet, we have aimed to provide as accurate as
possible classification in Table~\ref{tab:top}, in order to allow for
as fair as possible comparison of the two methods' NCFs under varying
sizes of network neighborhoods.

\begin{table}[h!]
\centering
\begin{tabular}{| l | l | l | }
\hline
Neighborhood size & Graphlet size & $k$-hop neighborhood \\
& (used by MI-GRAAL's NCF) & (used by GHOST's NCF) \\
\hline
T1 & 2-node graphlets & 1-hop neighborhood \\
\hline
T2 & 2-3-node graphlets & 2-hop neighborhood \\
\hline
T3 & 2-4-node graphlets & 3-hop neighborhood \\
\hline
T4 & 2-5-node graphlets & 4-hop neighborhood \\
\hline
\end{tabular} \\
\vspace{0.1cm}
\caption{The four neighborhood sizes that we vary within each aligner.}
\label{tab:top}
\end{table}

\noindent{\textbf{Implementation details.}} 
Types of scores that MI-GRAAL and GHOST take in as input are
different: MI-GRAAL looks at node similarities (the higher the score,
the more similar the nodes), while GHOST looks at node distances (the
lower the score, the more similar the nodes). We carefully take this
into account to allow for fair method comparison. For example, to
ensure that neither NCF has an advantage due to the format of the
scores, we normalize all scores. That is, node similarity scores used
in MI-GRAAL can exceed the value 1, while no scores generated by GHOST
are greater than one. To make the two sets of scores comparable, we
scale MI-GRAAL's node similarity scores to the [0-1] range by dividing
each of the scores by the maximum similarity score. Then, when we
give MI-GRAAL's NCF as input into GHOST's AS, because GHOST deals with
distances rather than similarities, we take one minus MI-GRAAL's
normalized NCF and then plug in the resulting node scores into
GHOST's AS. We do the analogous procedure to convert GHOST's node
distance scores into similarity scores when plugging GHOST's NCF into
MI-GRAAL's AS, which deals with similarities rather than distances.

Further, MI-GRAAL's NCF returns all pairwise node similarity scores
between two networks. However, GHOST's NCF returns only a subset of
the lowest of all pairwise node distance scores, depending on the
network size. To complete the GHOST's pairwise node score matrix and
thus allow for its to be given as input into MI-GRAAL's AS, we assign
a score equal to the highest distance score returned by GHOST to all
node pairs for which GHOST did not return a distance score.

Finally, current implementation of MI-GRAAL's AS does not function
properly when a large pairwise node similarity matrix is plugged into
it. Thus, MI-GRAAL's AS has had difficulty aligning two largest
networks from our study, those of fly and human. As a solution, we
create a matrix that contains only the top 21 million node similarity
scores of the original entire node similarity matrix, this being the
maximum that our computational resources would process. With this
adjustment, we are successfully able to generate all fly-human
alignments.

\subsection{Network alignment quality measures}\label{quality_measures}

We use well established network alignment quality measures
\cite{MilenkovicACMBCB2013,Faisal2014a,MAGNA}. Let $G_1(V_1,E_1)$ and
$G_2(V_2,E_2)$ be two graphs such that $|V_1| \leq |V_2|$. An
alignment of $G_1$ to $G_2$ is a total injective function $f : V_1
\rightarrow V_2$; every element of $V_1$ is matched uniquely with an
element of $V_2$. Let us denote by $E'_2$ the set of edges from $G_2$
that exist between nodes in $G_2$ that are aligned by $f$ to nodes in
$G_1$.

\noindent\textbf{Topological evaluation.} We use five
measures of topological alignment quality:

\begin{enumerate}
\item \emph{Node correctness (NC):} If $h : V_1 \rightarrow V_2$ is
the correct ground truth node mapping between $G_1$ and $G_2$ (when
such mapping is known), then NC of alignment $f$ is: $NC =
\frac{|\{u \in V_1 : h(u) = f(u) \}|}{|V_1|} \times 100\%$
\cite{GRAAL}. This measure can be computed only for alignments of
the synthetic network set with known ground truth node mapping
(Section \ref{data}). All remaining measures (listed below) can be
computed for the real network set with unknown node mapping as well.

\item \emph{Edge correctness (EC):} EC is the percentage of edges from
$G_1$, the smaller network (in terms of the number of nodes), which
are aligned to edges from $G_2$, the larger network \cite{GRAAL}.
Formally, $EC = \frac{|E_{1} \cap E'_2|}{|E_{1}|}\times{100\%}, $
where the numerator is the number of ``conserved'' edges, i.e., edges
that are aligned under the given node mapping. The larger the EC score, the
better the alignment.

\item \emph{Induced conserved structure (ICS)}: $ICS = \frac{|E_{1}
\cap E'_2|}{|E'_2|}\times{100\%}.$ EC might fail to differentiate
between alignments that one might intuitively consider to be of
different topological quality \cite{GHOST}, since it is defined with
respect to edges in $E_1$. For example, aligning a $k$-node cycle in
$G_1$ to a $k$-node cycle in $G_2$ would result in the same EC as
aligning a $k$-node cycle in $G_1$ to a $k$-node clique (complete
graph) in $G_2$. Clearly, the former is intuitively a better
alignment than the latter, since no edges that exist between the $k$
nodes in $G_2$ are left unaligned in the first case, whereas many
edges are left unaligned in the second case. Since ICS is defined
with respect to edges in $E'_2$, it would have the maximum value of
100\% when aligning a $k$-node cycle to a $k$-node cycle, and it
would have a lower value when aligning a $k$-node cycle to a
$k$-node clique \cite{MAGNA}. The larger the ICS, the better.

\item \emph{Symmetric substructure score (S$^3$)}: EC penalizes the
alignment for having misaligned edges in the smaller network. ICS
penalizes the alignment for having misaligned edges in the larger
network. S$^3$ on the other hand, aims to improve upon EC and ICS by
penalizing for misaligned edges in both the smaller and larger
network. S$^3 = \frac{|E_{1} \cap E'_2|}{|E_1| +|E'_2| - |E_{1} \cap
E'_2|}\times{100\%}.$ For details, see the original publication
\cite{MAGNA}.

\item The size of the \emph{largest connected common subgraph (LCCS)}
\cite{GRAAL}, which we use for the following reason. Of two
alignments with similar EC, ICS, or S$^3$ scores, one could expose
large, contiguous, and topologically complex regions of network
similarity, while the other could fail to do so. Thus, in addition
to counting aligned edges or nodes that participate in the aligned
edges, it is important that the aligned edges cluster together to
form large connected subgraphs rather than being isolated. Hence, we
define a connected common subgraph (CCS) as a connected subgraph
(not necessarily induced) that appears in both networks
\cite{HGRAAL}. We measure the size of the largest CCS (LCCS) in
terms of the number of nodes as well as edges. Namely, we compute
the LCCS score as in our recent work \cite{MAGNA}. First, we count
$N$, the percentage of nodes from $G_1$ that are in the LCCS. Then,
we count $E$, the percentage of edges that are in the LCCS out of
all edges that could have been aligned between the nodes in the
LCCS. That is, $E$ is the minimum of the number of edges in the
subgraph of $G_1$ that is induced on the nodes from the LCCS, and
the number of edges in the subgraph of $G_2$ that is induced on the
nodes from the LCCS \cite{MAGNA}. Finally, we compute their
geometric mean as $\sqrt(N \times E)$, in order to penalize
alignments that have small $N$ or small $E$. Large values of this
final LCCS score are desirable.

\end{enumerate}

\noindent\textbf{Biological evaluation.} Only alignments in which many
aligned node pairs perform the same function should be used to
transfer function from annotated parts of one network to unannotated
parts of another network \cite{MAGNA}. Hence, we measure Gene Ontology
(GO) \cite{Go00} enrichment of aligned proteins pairs, i.e., the
percentage of protein pairs in which the two proteins \emph{share} at
least one GO term, out of all aligned protein pairs in which both
proteins are annotated with at least one GO term. We refer to this
percentage as \emph{GO correctness (GO)}. We do this with respect to
complete GO annotation data, independent of GO evidence code. Also,
since many GO annotations have been obtained via sequence comparison,
and since some of the aligners use sequence information, we repeat the
analysis considering only GO annotations with experimental evidence
codes, in order to avoid the circular argument. In this case, we refer
to GO correctness as \emph{experimental GO correctness (EXP)}. The
higher the GO and EXP values, the better \cite{MAGNA}.

\section{Results and discussion} \label{results}

We aim to answer the following three main questions in the context of
network alignment: {\bf 1)} What is the best NCF and the best AS, and
is there perhaps a combination of one existing method's NCF and
another existing method's AS that is the superior aligner (Section
\ref{best})? {\bf2)} How much sequence versus topological information
to use within NCF (Section \ref{SvT})? {\bf3)} How large the size of
network neighborhoods of compared nodes to consider within NCF
(Section \ref{size})? In addition, we comment on relationships between
different alignment quality measures (Section \ref{correlate}).
Finally, we conclude in Section~\ref{conclusion}.

\subsection{What is the best NCF and the best AS?} \label{best}

By comparing M-M and G-M aligners, we can fairly compare the two NCFs
under the same (MI-GRAAL's) AS. Also, by comparing G-M and G-G, we can
fairly compare the two ASs under the same (GHOST's) NCF. See Section
\ref{generate} for details.

\subsubsection{Synthetic networks with known node mapping} \label{best:known}

Overall, GHOST's NCF is slightly superior to that of MI-GRAAL (Fig.
\ref{fig:align_bars} (a)-(b)). Also, GHOST's AS is superior to
MI-GRAAL's AS (Fig. \ref{fig:align_bars} (a)-(b)). However, these
findings are based on \emph{all} alignments (with known node mapping)
for all values of $\alpha$, all neighborhood sizes, and all measures
of alignment quality combined (Section \ref{generate}), which might
not be fair. Thus, in Fig.~\ref{fig:best_yeast} (a)-(c), for each
aligner, for each alignment quality measure, we show results for
the \emph{best} alignments over all values of $\alpha$ and all
neighborhood sizes, for three out of all five network pairs (for the
remaining network pairs, see the Supplement). Now, the
general trend (and especially with respect to NC as the most accurate
ground truth measure of alignment quality) is that the best scores for
M-M are either comparable or superior to those of G-M, indicating
slight superiority of MI-GRAAL's NCF over GHOST's. Nonetheless, G-G
still always outperforms G-M, indicating superiority of GHOST's AS
over MI-GRAAL's AS.

It is possible to break down the above results and study how the
ranking of the different NCFs and ASs changes with the change in the
value of $\alpha$, which corresponds to the amount of topological
similarity information used within NCF (see 
the Supplement). In
general, MI-GRAAL's NCF is comparable to GHOST's NCF across all
$\alpha$ values, as M-M and G-M scores are similar. On the other hand,
GHOST's AS shows superiority over MI-GRAAL's AS, as G-G consistently
results in higher scores than G-M. We note that we show that the value
of $\alpha$ does not greatly affect alignment quality
(Section~\ref{SvT}).

It is also possible to break down the above results even further and
study how the ranking of the different NCFs and ASs changes with the
change in the neighborhood size that is considered within NCF
(see 
the Supplement). In general, for the smaller neighborhood
sizes (T1 and T2), GHOST's NCF generally produces comparable or
superior results to MI-GRAAL's NCF, as G-M scores are higher than M-M
scores. However, for the larger neighborhood sizes (T3 and T4),
MI-GRAAL's NCF is comparable or superior to GHOST's NCF. And because
we show that the larger neighborhood sizes (T3 and T4) are overall
superior (Section \ref{size}), this means that overall MI-GRAAL's NCF
is comparable to or superior to GHOST's NCF. On the other hand, in
general, for all network sizes, GHOST's AS consistently outperforms
MI-GRAAL's AS, as G-G scores is typically higher than G-M scores.

\subsubsection{Real networks with unknown node mapping} \label{best:unknown}

\begin{figure*}[h]
\begin{center}
\includegraphics[width=3.5in]{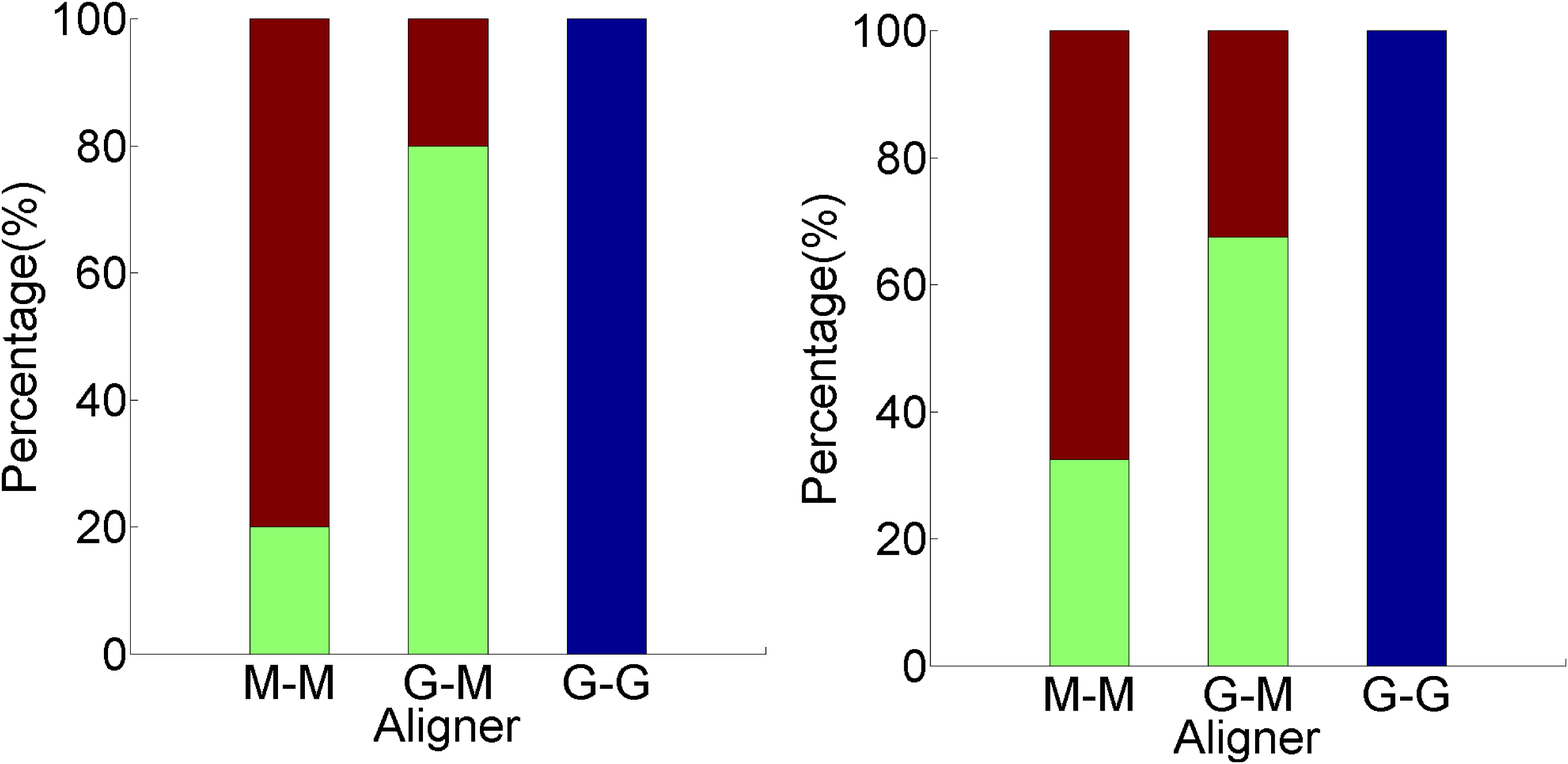}\\
\hspace{.4in}\textbf{(a)}\hspace{1.55in}\textbf{(b)}\\
\includegraphics[width=3.5in]{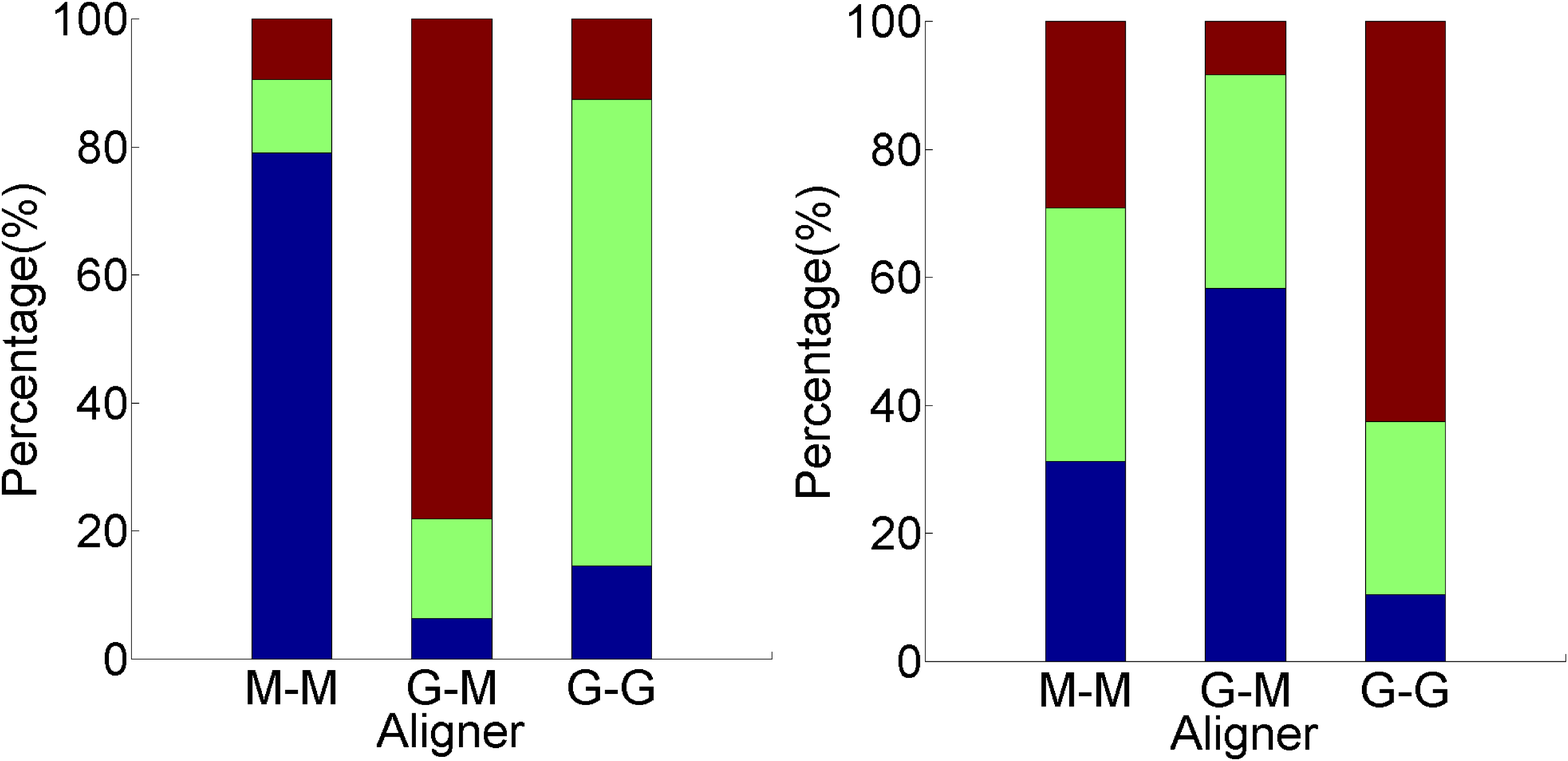}\\
\hspace{.4in} \textbf{(c)} \hspace{1.55in}\textbf{(d)}\\
\includegraphics[width=.8in]{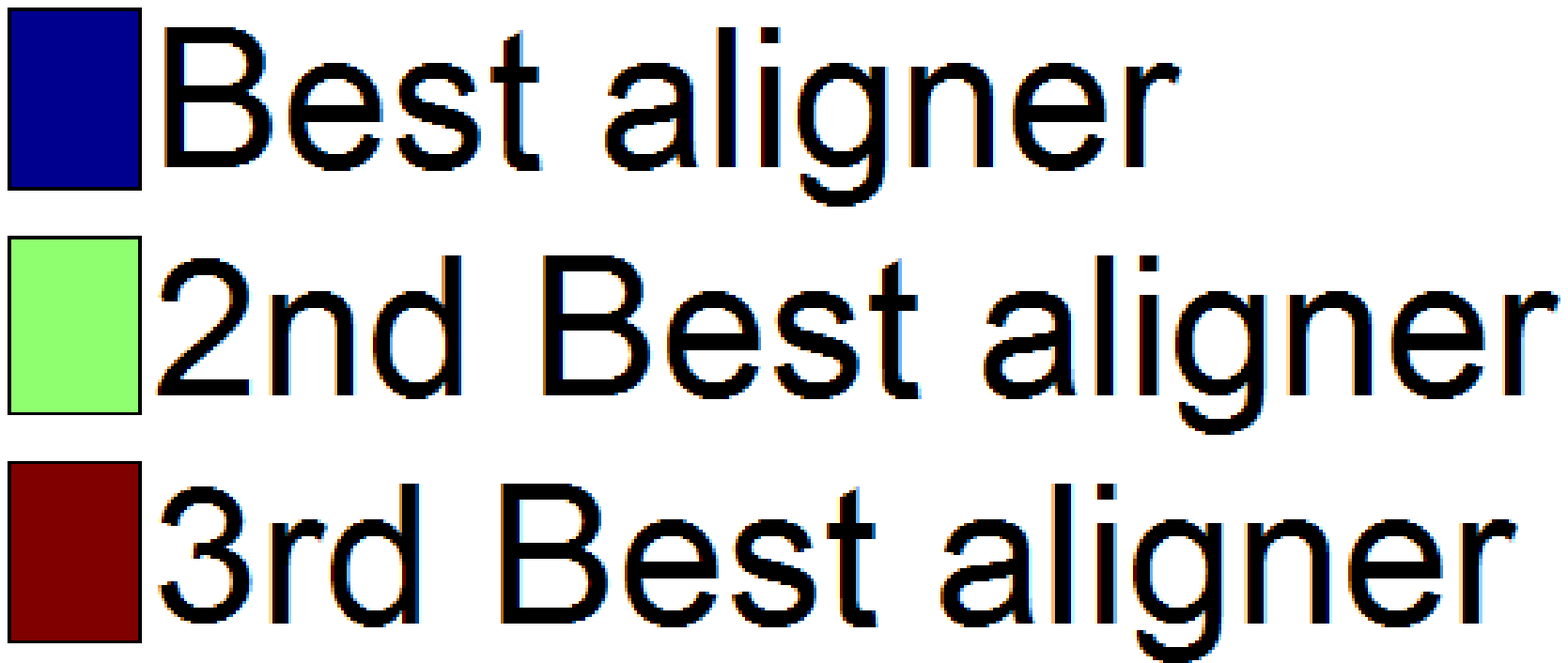}
\end{center}
\caption{The ranking of the three aligners (M-M, G-M, and G-G) over
\emph{all} alignments for all values of $\alpha$ and all
neighborhood sizes, with respect to: {\bf (a)} all topological
scores of all alignments with known ground truth node mapping, {\bf
(b)} all biological scores of alignments with known node mapping,
{\bf (c)} all topological scores of alignments with unknown node
mapping, and {\bf (d)} all biological scores of alignments with
unknown node mapping.}
\label{fig:align_bars}
\end{figure*}

Overall, unlike for the synthetic network data set with known node
mapping, on the real network data set with unknown mapping, MI-GRAAL's
NCF is now comparable or superior to that of GHOST (Fig.
\ref{fig:align_bars} (c)-(d)). Further, MI-GRAAL's AS is now
comparable or superior to GHOST's AS (Fig. \ref{fig:align_bars}
(c)-(d)). We confirm these findings even when we limit from \emph{all}
alignments (Fig. \ref{fig:align_bars} (c)-(d)) to the \emph{best}
alignments only (just as above) (Fig.~\ref{fig:best_yeast} (d)-(f))
(see 
the Supplement).

When zooming into the results further to observe the effect of the
$\alpha$ parameter, in general, for all values of $\alpha$, MI-GRAAL's
NCF is comparable or superior to GHOST's NCF and MI-GRAAL's AS is
comparable to GHOST AS across all values of $\alpha$ (see 
the Supplement).

The same holds independent on the neighborhood size that is
considered within NCF (see 
the Supplement).

\begin{figure*}[h!]
\begin{center}
\includegraphics[width=4.9in]{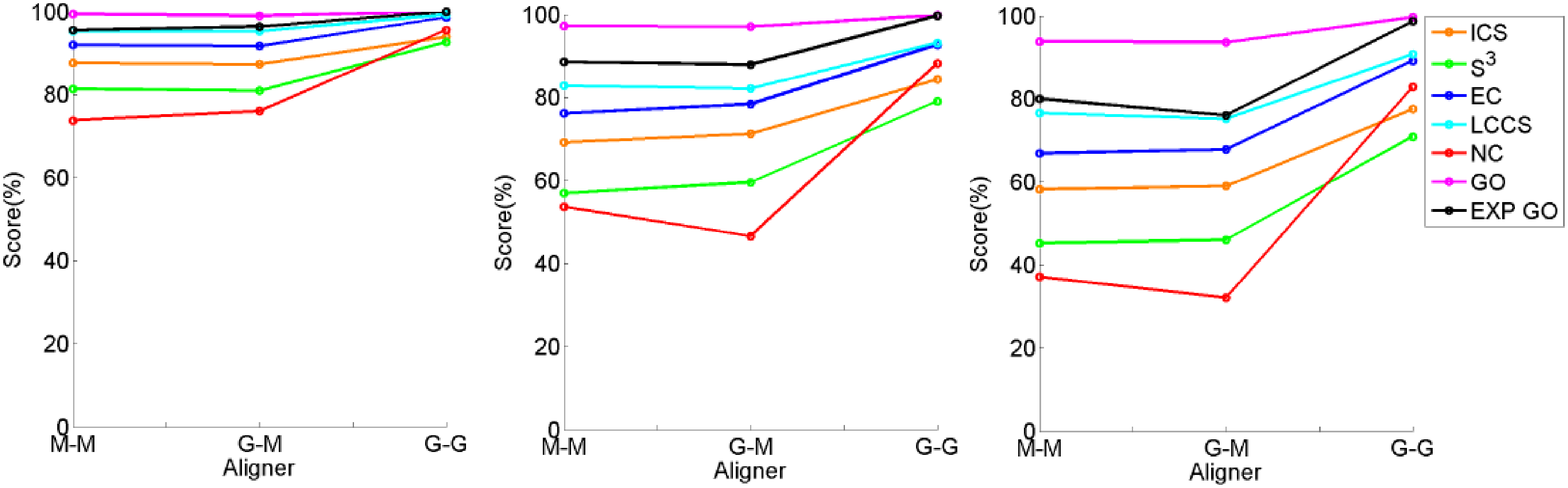} \\
\hspace{-.1in}\textbf{(a)}\hspace{1.4in}\textbf{(b)} \hspace{1.25in} \textbf{(c)}
\includegraphics[width=4.9in]{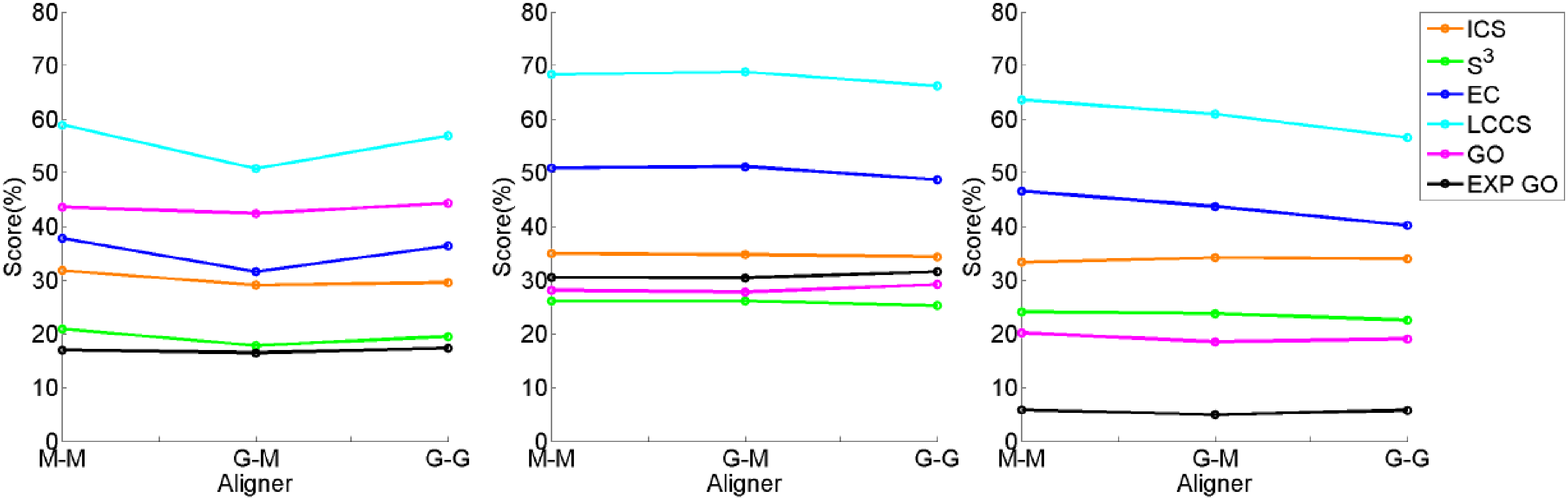} \\
\hspace{-.1in}\textbf{(d)}\hspace{1.4in}\textbf{(e)} \hspace{1.25in} \textbf{(f)}
\end{center}
\caption{Alignment quality results of the three aligners (M-M, G-M,
and G-G) for \emph{best} alignments over all values of $\alpha$ and
all neighborhood sizes, for (a)-(c) three network pairs with known
node mapping (yeast-yeast 5\%, yeast-yeast 10\%, and yeast-yeast
15\%, respectively) and (d)-(e) three network pairs with unknown
mapping (human-yeast, human-worm, and worm-yeast, respectively). For
equivalent results for the remaining network pairs, see 
the Supplement.}
\label{fig:best_yeast}
\end{figure*}

\subsubsection{Summary}

Which NCF or AS is the best \emph{overall} is not easy to determine,
as the results are data-dependent. But when we limit analyses of each
aligner to the \emph{best} alignments over all parameters, M-M is
comparable or superior to G-M, indicating that MI-GRAAL's NCF is
better than GHOST's NCF, while the performance of G-M versus G-G,
i.e., of MI-GRAAL's AS versus GHOST's AS, is still data-dependent.

Therefore, the graphlet-based measure of topological node similarity
\cite{Milenkovic2008} that MI-GRAAL uses (along with many other
network aligners \cite{GRAAL,HGRAAL,MilenkovicACMBCB2013,Faisal2014a}
or even network clustering methods
\cite{Milenkovic2008,MMGP_Roy_Soc_09,Solava2012}) remains the
state-of-the-art, even when compared to the spectral signature-based
node similarity measure that GHOST uses (and especially compared to
PageRank-based node similarity measure that aligners from the IsoRank
family use, as we already showed in our recent study
\cite{MilenkovicACMBCB2013,Faisal2014a}).

Our results indicate that the slight superiority of GHOST (i.e., G-G)
over MI-GRAAL (i.e., M-M) that was claimed in the original GHOST
publication \cite{GHOST} seems to come from GHOST's AS and \emph{not}
its NCF, which is not surprising, since GHOST's AS solves quadratic
assignment problem whereas MI-GRAAL's AS deals only with linear
assignment problem. Further, our results indicate that the combination
of MI-GRAAL's NCF and GHOST's AS (i.e., M-G) could be a new aligner
that is superior to the existing MI-GRAAL (i.e., M-M) and GHOST (ie.,
G-G) aligners on at least some data sets. Unfortunately, explicitly
testing this is not possible with the current implementation of GHOST,
as per our conversation with the authors of GHOST, the current
implementation is too complex to modify to allow for plugging
MI-GRAAL's (or any other method's) NCF into GHOST's AS.

\subsection{The amount of sequence versus topological information
within NCF?} \label{SvT}

Recall that we vary the amount of topological node similarity
information within NCF with the $\alpha$ parameter (where $\alpha$ of
0 means that no topology information is used, i.e., that only sequence
information is used, whereas $\alpha$ of 1 means that only topology
information is used; Section \ref{generate}). Here, we study the
effect of the $\alpha$ parameter on alignment quality.

\begin{figure*}[t!]
\begin{center}
\includegraphics[width=5in]{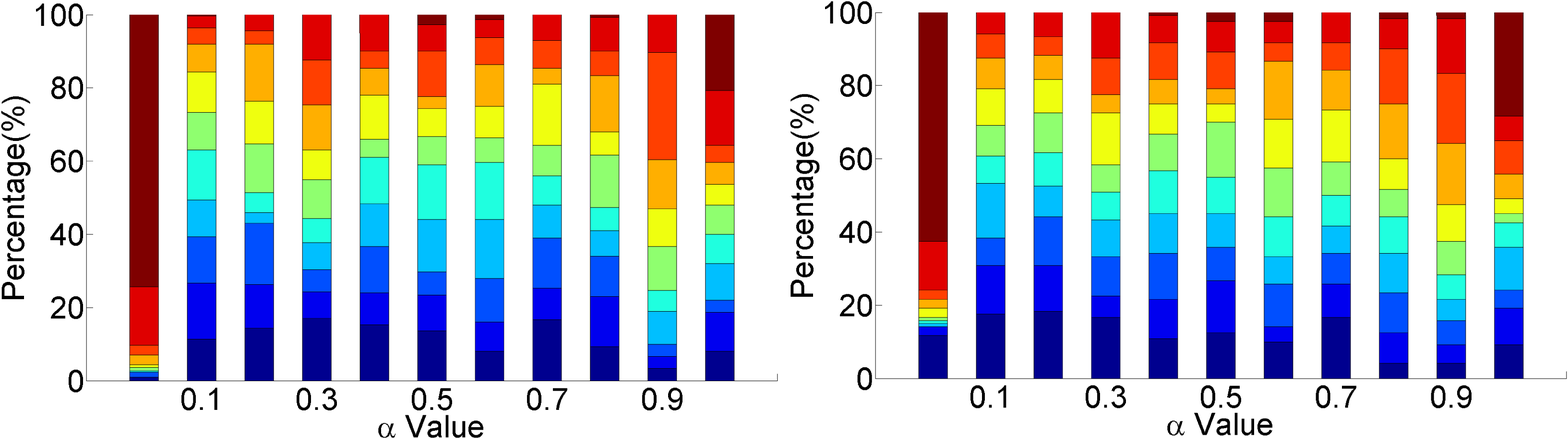}\\
\hspace{.6in}\textbf{(a)}\hspace{2.4in}\textbf{(b)} 
\includegraphics[width=5in]{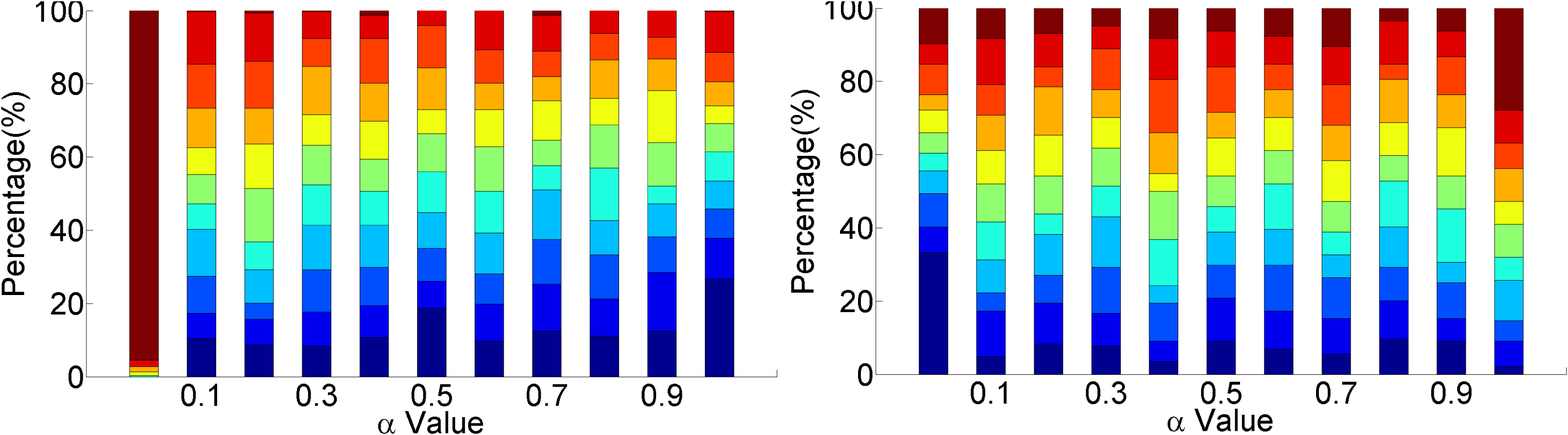}\\
\hspace{.6in}\textbf{(c)}\hspace{2.4in}\textbf{(d)} \\
\includegraphics[width=1.1in]{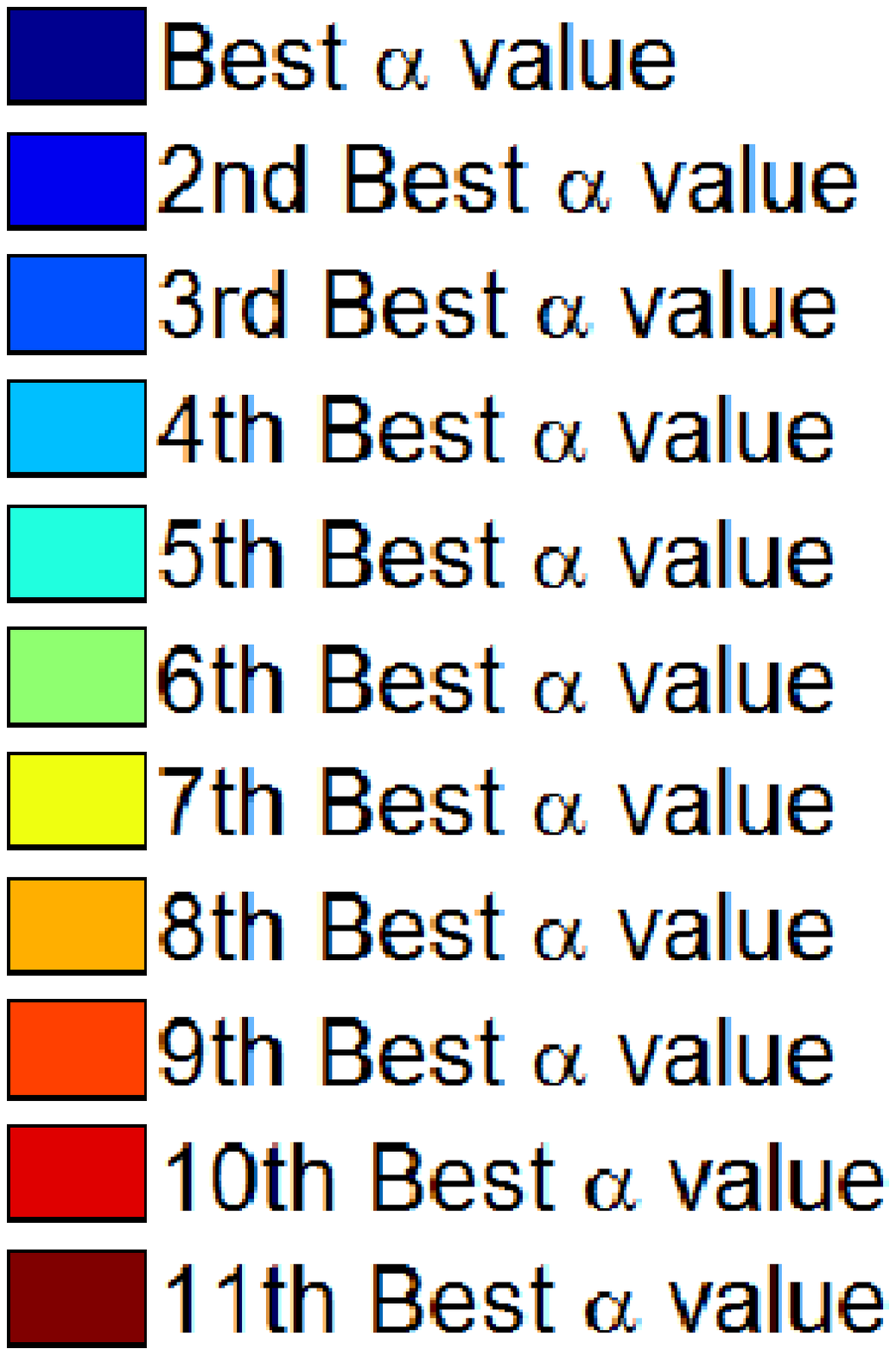}
\end{center}
\caption{The ranking of the 11 values of $\alpha$ (from 0 to 1 in
increments in 0.1) over \emph{all} alignments for all aligners and
all neighborhood sizes, with respect to: {\bf (a)} all topological
scores of alignments with known ground truth node mapping, {\bf (b)}
all biological scores of alignments with known node mapping, {\bf
(c)} all topological scores of alignments with unknown node
mapping, and {\bf (d)} all biological scores of alignments with
unknown node mapping.}
\label{fig:alpha}
\end{figure*}

\subsubsection{Synthetic networks with known node mapping}\label{SvT:known}

Overall, the value of $\alpha$ does not affect alignment quality, as
long as some amount of topological information is used. That is, only
$\alpha=0.0$ results in completely inferior alignments, especially
with respect to topological alignment quality, whereas all other
values of alpha are more-less comparable
(Fig.~\ref{fig:alpha} (a)-(b)). It is expected that the larger the
value of $\alpha$, i.e., the more of topological information is used
within NCF, the better the topological alignment quality. Again, this
is exactly what we observe (Fig.~\ref{fig:alpha} (a)). It is also
expected that that the smaller the value of $\alpha$, i.e., the more
of sequence information is used within NCF, the better the biological
alignment quality. Surprisingly, this is not what we observe
(Fig.~\ref{fig:alpha} (b)): larger values of $\alpha$ (e.g., 0.7)
result in more of high-quality alignments than $\alpha=0$.

\begin{figure}[h]
\begin{center}
\includegraphics[width=4.9in]{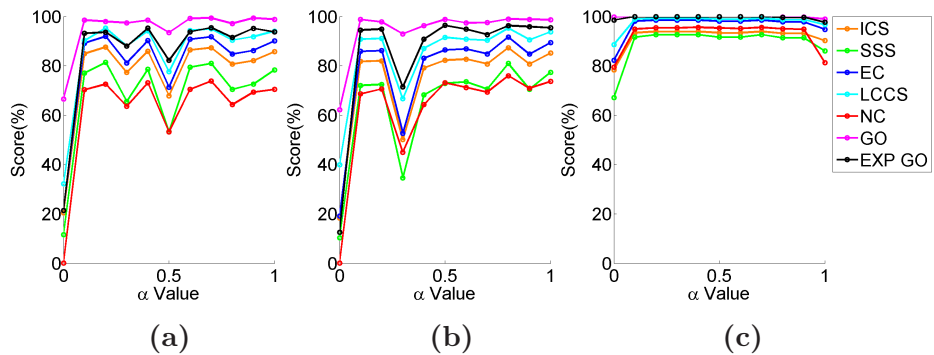}\\
\hspace{-.2in}{\bf (a)} \hspace{1.2in}{\bf (b)} \hspace{1.2in} {\bf (c)}
\end{center}
\caption{Detailed illustration of the effect of the $\alpha$ parameter for 
{\bf (a)} M-M, {\bf (b)} G-M, and {\bf (c)} G-G aligners. In
particular, results are shown for the yeast-yeast 5\% alignment and
for the neighborhood size T4. For other network pairs and other
neighborhood sizes, see 
the Supplement for synthetic network
data and see 
the Supplement for real-world PPI network data.}
\label{fig:seq_example}
\end{figure}

When zooming into the results further to observe the effect of the
aligner, in general, we see the same trends as above independent of
the aligner (see 
the Supplement). Namely, the results from
Fig.~\ref{fig:alpha} (a)-(b) hold independent on which NCF or AS is
used. Further, there is no difference in the results across the two
NCFs (Fig.~\ref{fig:seq_example} (a) and (b)). There is only a minor
difference in the results across the two ASs, in the sense that the
results are somewhat more stable across different $\alpha$s for
GHOST's AS than for MI-GRAAL's AS (Fig.~\ref{fig:seq_example} (b) and
(c)). Also, GHOST's AS suggests that in addition to not using
$\alpha=0$ (i.e., sequence alone), one should not use $\alpha=1$
either (i.e., topology alone); but other than that, the choice of
$\alpha$ still has no major effect (Fig.~\ref{fig:seq_example} (c)).

When zooming into the results from Fig.~\ref{fig:alpha} (a)-(b)
further to observe the effect of the neighborhood size, we see
that the results hold independent of the neighborhood size
(see 
the Supplement).

\subsubsection{Real networks with unknown node mapping}\label{SvT:unknown}

The results that we observe for the synthetic networks in general hold
for this network set as well. Namely, $\alpha=0$ results in the worst
topological alignment quality, while the other $\alpha$ values are
somewhat comparable, with a slight dominance of the larger values, as
expected (Fig.~\ref{fig:alpha} (c)). Interestingly, for this network
set, the lowest value of $\alpha=0$ results in the most of
highest-scoring alignments with respect to biological alignment
quality; yet, even the largest $\alpha$s often lead to good alignments
with respect to biological alignment quality (Fig.~\ref{fig:alpha}
(d)).

When zooming into the results further to observe the effect of the
aligner, as with synthetic networks, the general results from
Fig.~\ref{fig:alpha} (c) and (d) hold independent of the aligner for
real networks as well (see 
the Supplement). However, unlike for
synthetic networks, for real networks we now see result stability
across all NCFs and all ASs, and not just for GHOST's AS. Also, now
GHOST's AS no longer suggests that $\alpha=1$ should not be used.

When zooming into the results from Fig.~\ref{fig:alpha} (c) and (d)
further to observe the effect of the neighborhood size, just as with
the synthetic networks, we again see that the results hold independent
of the neighborhood size (see 
the Supplement). 

\subsubsection{Summary}\label{SvT:summary}

Overall, at least some amount of topological information should be
included within NCF, as this results in good topological as well as
biological alignment quality. While $\alpha=0.0$ may (but does not
always) result in biologically high-quality alignments, in every case
it fails to produce topologically superior results. Thus, $\alpha=0.0$
should not be used.

\subsection{The size of nodes' neighborhoods within NCF?} \label{size}

Intuitively, one would expect that the increase in the size of nodes'
network neighborhoods within NCF (i.e., in the amount of network
topology) would result in higher-quality alignments. However, this
assumption has not been tested to date. Instead, the existing methods
blindly use the largest neighborhood size that is allowed by available
computational resources (that is, MI-GRAAL uses all 2-5-node
graphlets, whereas GHOST uses $k=4$; Section \ref{generate}). Thus,
within each aligner, we vary the neighborhood size from T1 to T4
(Table~\ref{tab:top}) to systematically evaluate the effect of this
parameter.

\subsubsection{Synthetic networks with known node mapping} \label{size:known}

\begin{figure*}[t]
\begin{center}
\includegraphics[width=3.5in]{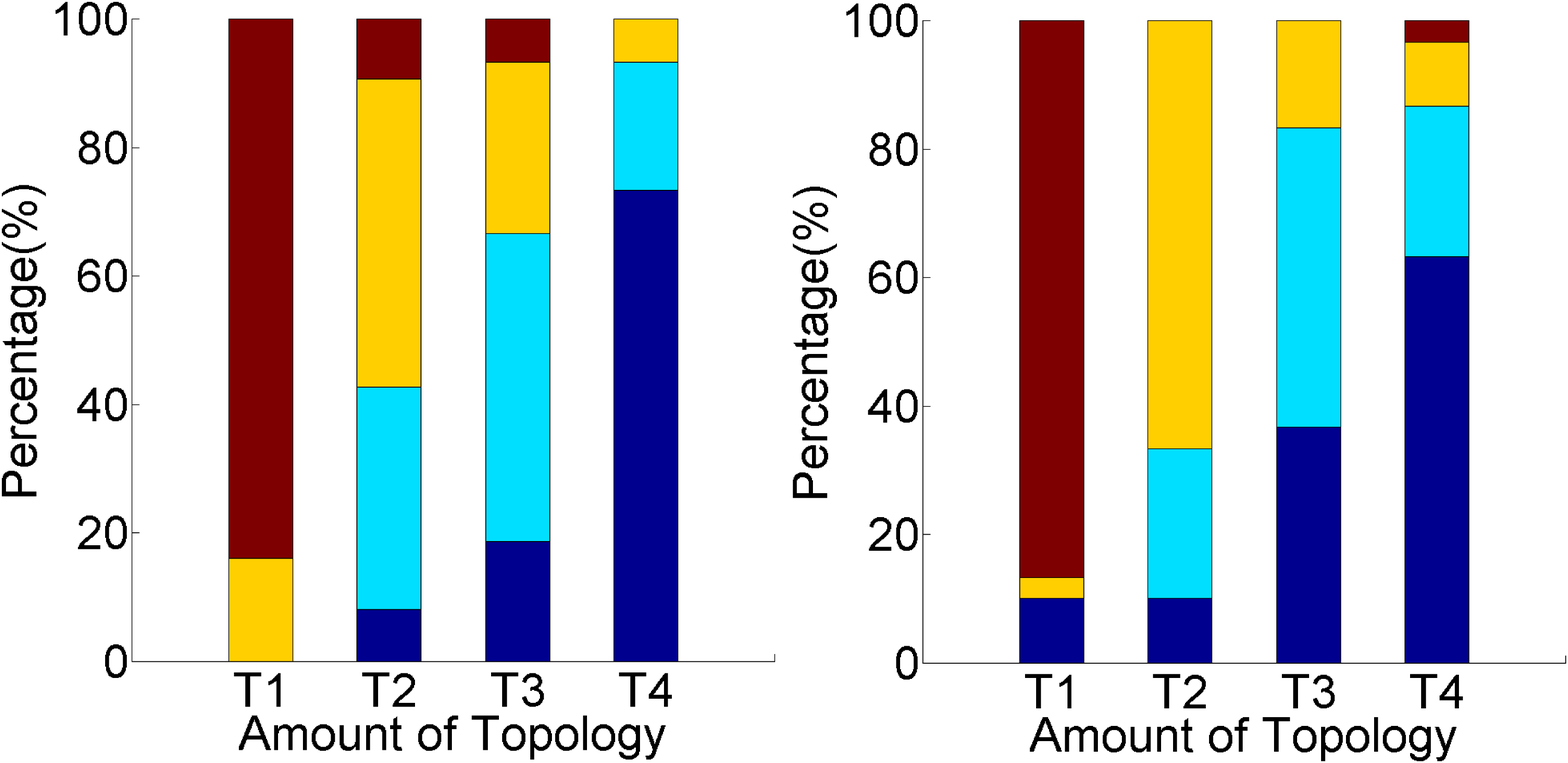}\\
\hspace{.4in}\textbf{(a)}\hspace{1.55in}\textbf{(b)}\\
\includegraphics[width=3.5in]{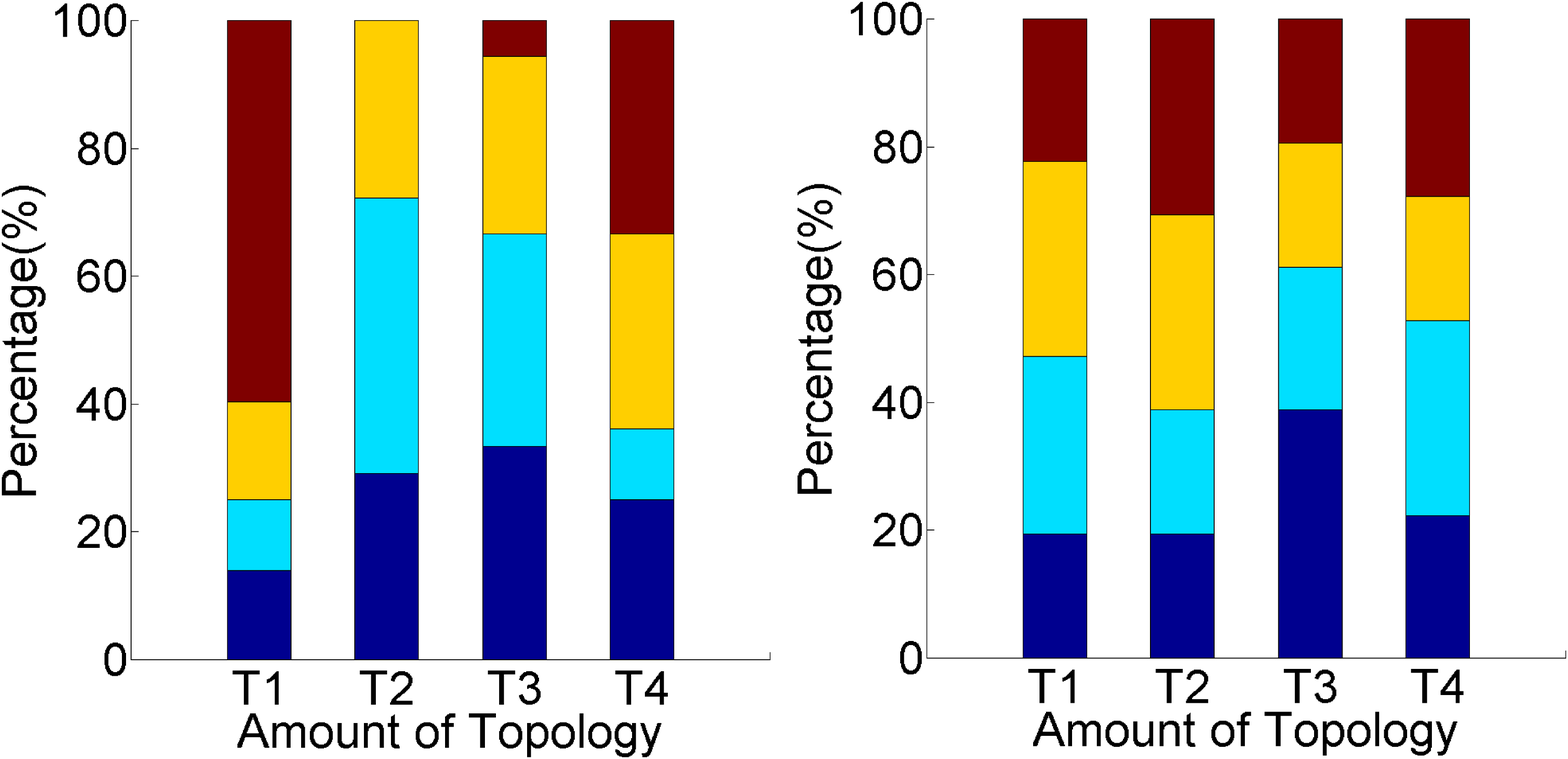}\\
\hspace{.4in}\textbf{(c)}\hspace{1.55in}\textbf{(d)} \\
\includegraphics[width=.8in]{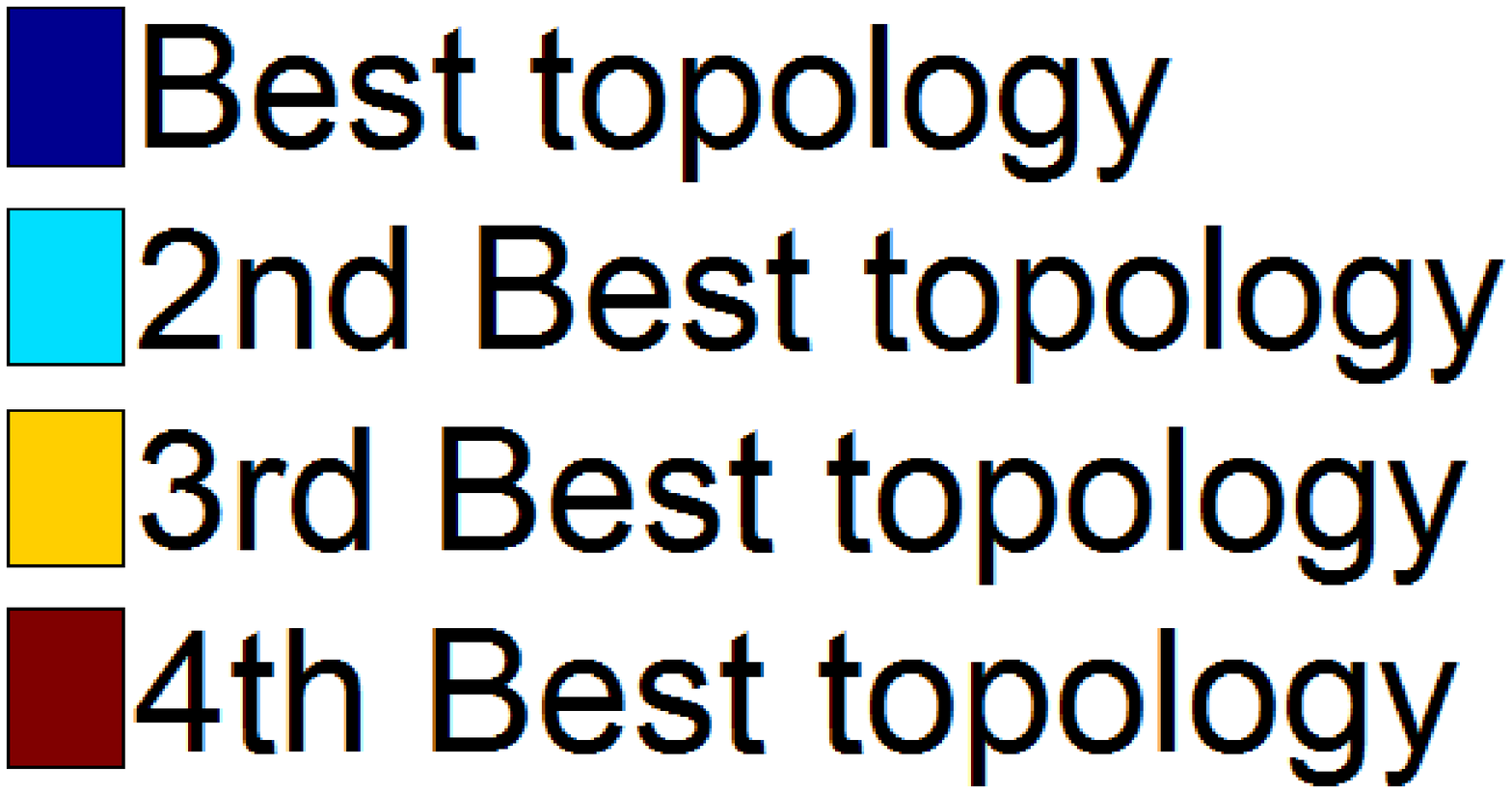}
\end{center}
\caption{The ranking of the four neighborhood sizes (T1-T4) over
\emph{all} alignments for all aligners and all values of $\alpha$,
with respect to: {\bf (a)} all topological scores of alignments with
known ground truth node mapping, {\bf (b)} all biological scores of
alignments with known node mapping, {\bf (c)} all topological scores
of alignments with unknown node mapping, and {\bf (d)} all
biological scores of alignments with unknown node mapping.}
\label{fig:top_bars}
\end{figure*}

Overall, the larger the neighborhood size, the better the alignment
quality, even though all neighborhood sizes except T1 can in some
cases result in higher-quality alignments than any other neighborhood
size (Fig.~\ref{fig:top_bars} (a)-(b)). That is, for some values of
network alignment parameters, smaller neighborhoods can produce
higher-quality alignments than larger neighborhoods, which is a
surprising result.

When zooming into the results further to observe the effect of the
aligner, the general trends from Fig.~\ref{fig:top_bars} (a)-(b) still
hold independent of the aligner, but some fluctuations in the results
exist (see 
the Supplement). Namely, M-M generally prefers T3 and
T4 neighborhood sizes. G-M prefers T2 in addition to T3 and T4, where
T3 or T4 are actually inferior to T2 in some cases, depending on the
noise level. G-G performs well on of T1-T4, with a slight preference
of T3 or T4, depending on the noise level. See
Fig.~\ref{fig:top_example} (a) for an illustration.

When zooming into the results further to observe the effect of the
$\alpha$ parameter, general trends from Fig.~\ref{fig:top_bars}
(a)-(b) are overall the same for all values of $\alpha$ (see 
the Supplement). The only exception is $\alpha=0$, which should not be
used in the first place (Section \ref{SvT:summary}).

\subsubsection{Real networks with unknown node mapping} \label{size:unknown}

Unlike for the synthetic networks, the largest neighborhood size (T4)
is now not overly dominant over the smaller network sizes.
Specifically, for real network data set, it is T3 that is the most
dominant, followed by T4 and T2, which are tied, and followed by T1,
which is inferior (Fig.~\ref{fig:top_bars} (c) and (d)).

\begin{figure}[h]
\begin{center}
\textbf{(a)}\includegraphics[width=2.1in]{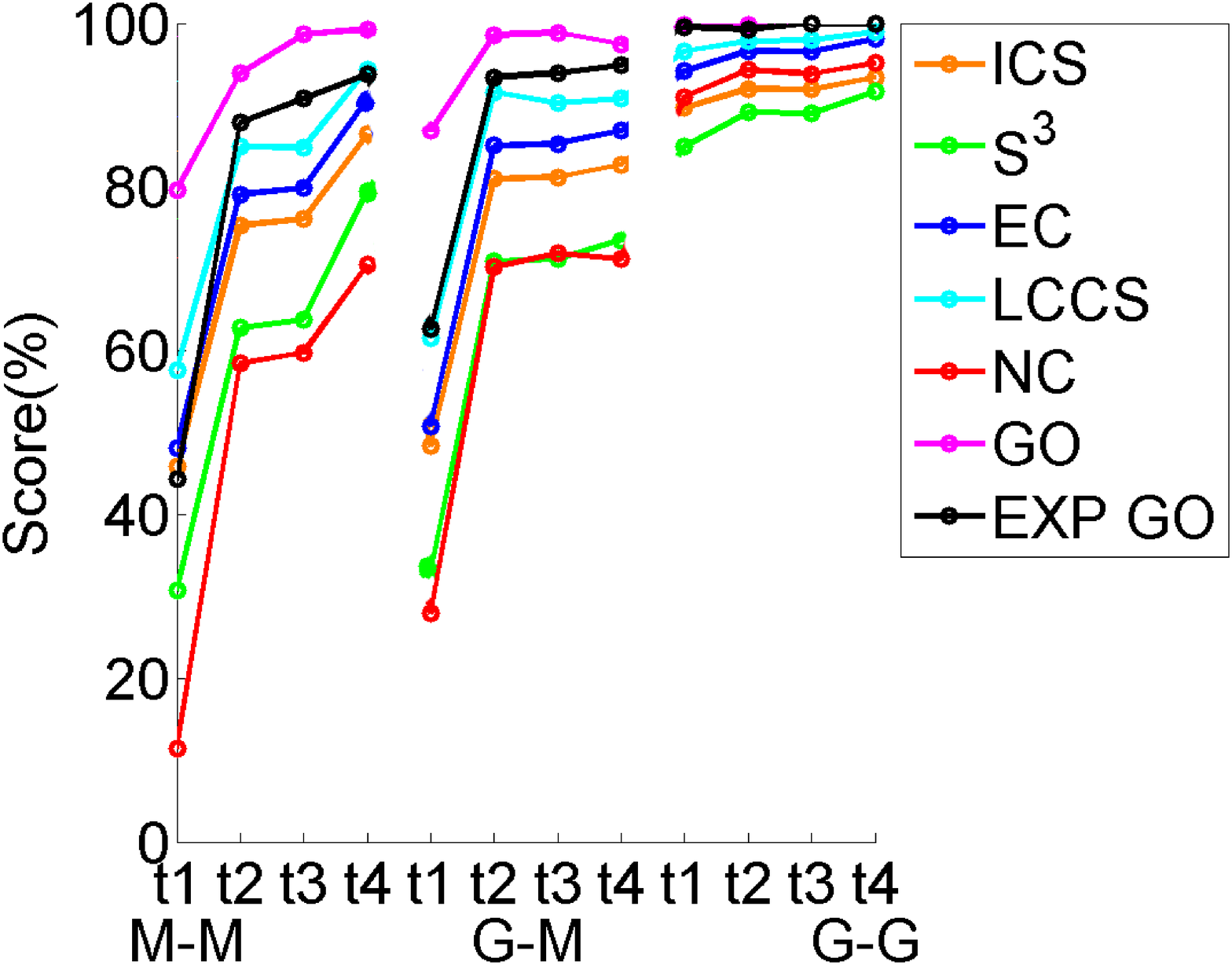}
\textbf{(b)}\includegraphics[width=2.1in]{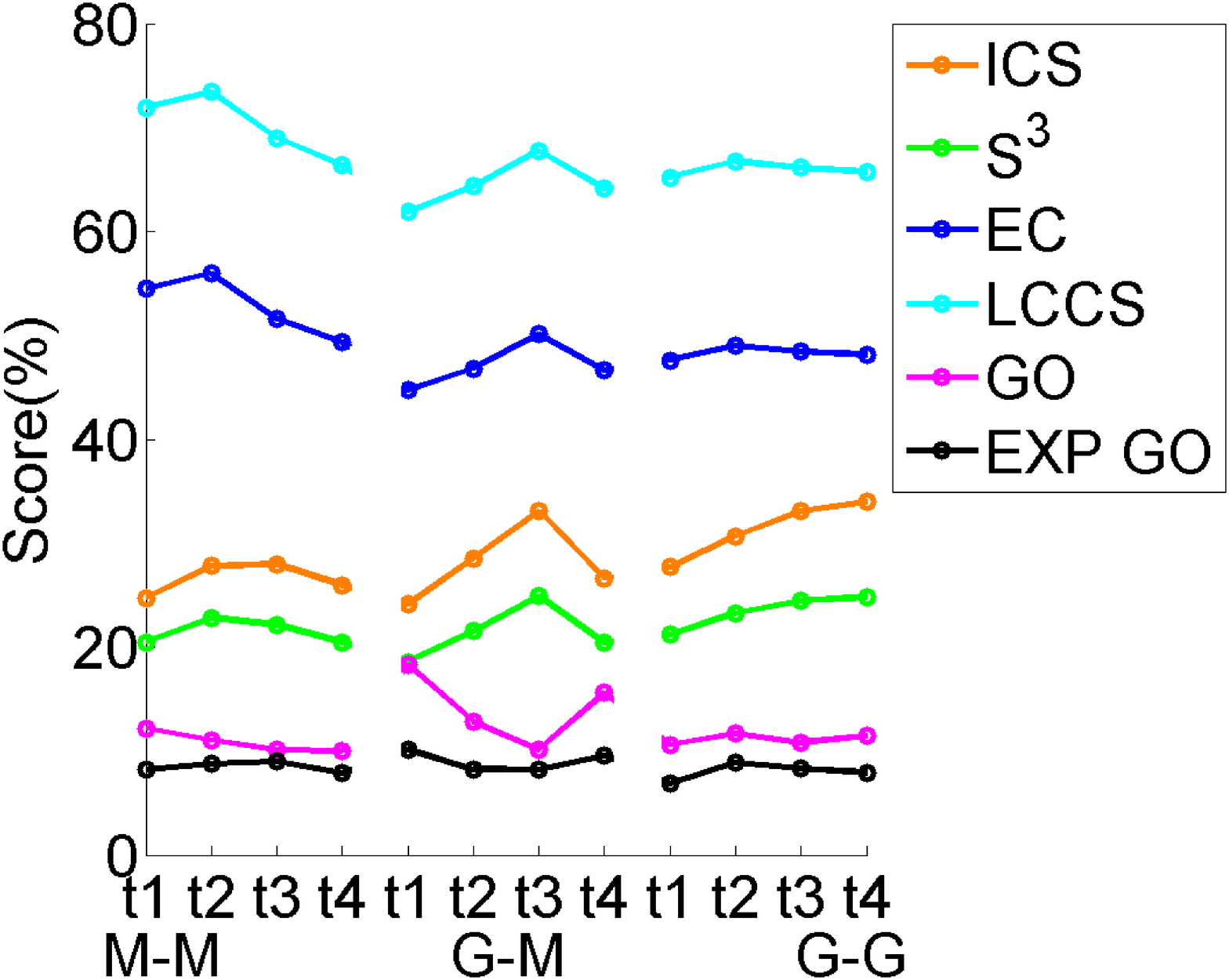}
\end{center}
\caption{Detailed illustration of the effect of the neighborhood size for 
{\bf (a)} synthetic and {\bf (b)} real network data. In particular,
results are shown for all three aligners, for the yeast-yeast 5\%
alignment at $\alpha=0.6$ in panel (a) and for the fly-worm
alignment at $\alpha=0.4$ in panel (b). For other network pairs and
other values of $\alpha$, see 
the Supplement.}
\label{fig:top_example}
\end{figure}

When zooming into the results further to observe the effect of the
aligner, we see that each aligner has an interesting behavior
(see 
the Supplement). Namely, M-M's and G-G's preference on the
neighborhood size is mainly dictated by the choice of species whose
networks are aligned. For G-M, in general, the larger
neighborhood sizes are preferred; in some cases, depending on the
species, G-M prefers T3 more than other neighborhood sizes. See
Fig.~\ref{fig:top_example} (b) for an illustration.

When zooming into the results further to observe the effect of the
$\alpha$ parameter, just as for synthetic networks, the results from
Fig.~\ref{fig:top_bars} (c) and (d) do not drastically change with the
change of $\alpha$ value (see 
the Supplement).

\subsubsection{Summary}

In general, the larger the neighborhood size within NCF, the higher
the alignment quality. However, it is not necessarily the case that the
largest neighborhood size always produces the best alignments nor that
it is always dominant to the smaller neighborhood sizes. This means
that slightly smaller neighborhood sizes (and T3 in particular) might
be desirable, as this could not only produce better alignments in
some cases but also decrease the computational complexity of the given
method.

\subsection{Relationships between different alignment quality
measures} \label{correlate}

We use a total of seven alignment quality measures: the ground truth
NC measure that can only be measured in alignments of synthetic
networks with known node mapping, four additional topological measures
(EC, ICS, S$^3$, and LCCS), and two biological measures (GO and EXP)
(Section~\ref{quality_measures}). Here, we briefly comment on the
relationship between the different measures.

\begin{figure}
\centering
\textbf{(a)}\includegraphics[width=2.1in]{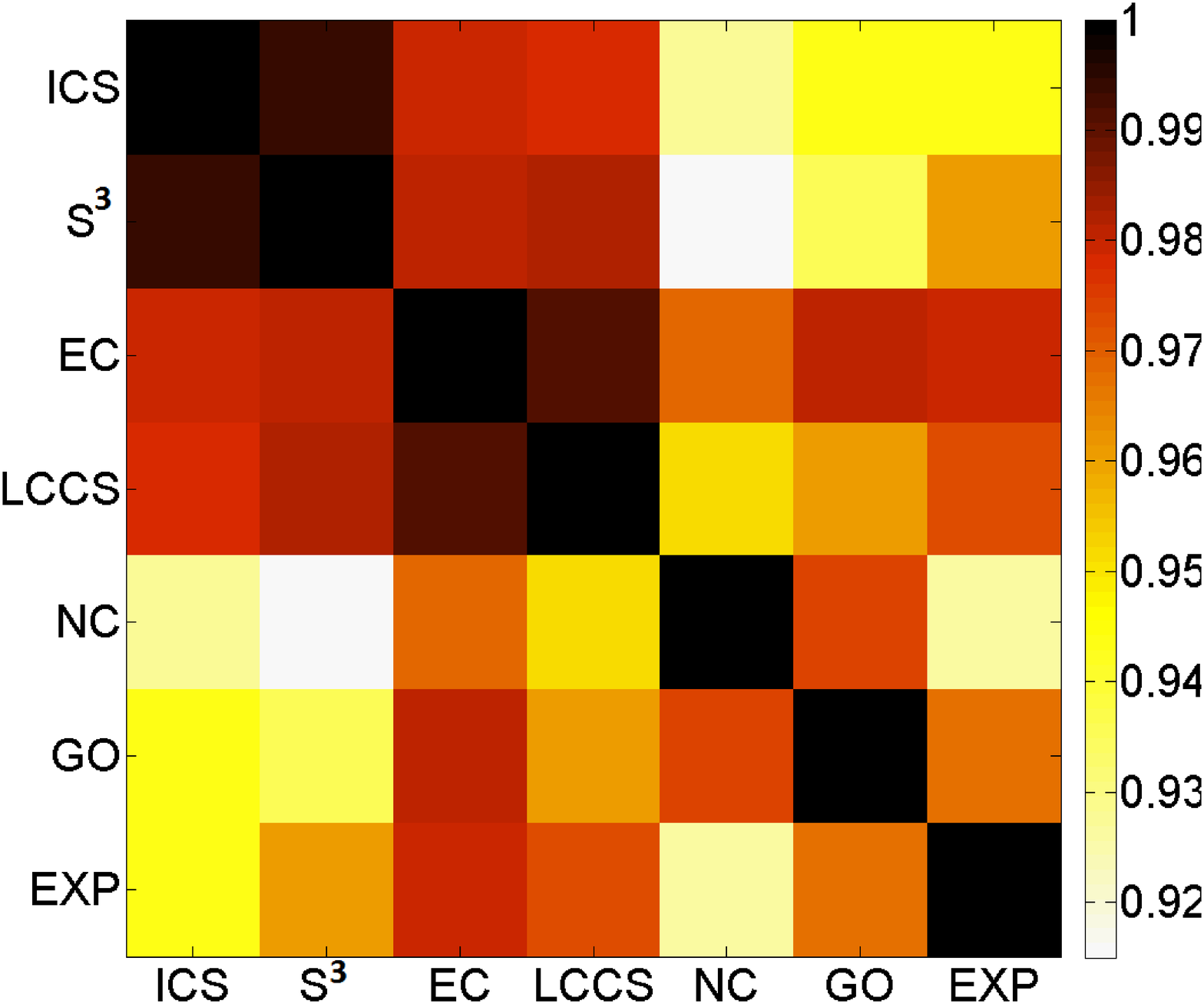}
\textbf{(b)}\includegraphics[width=2.1in]{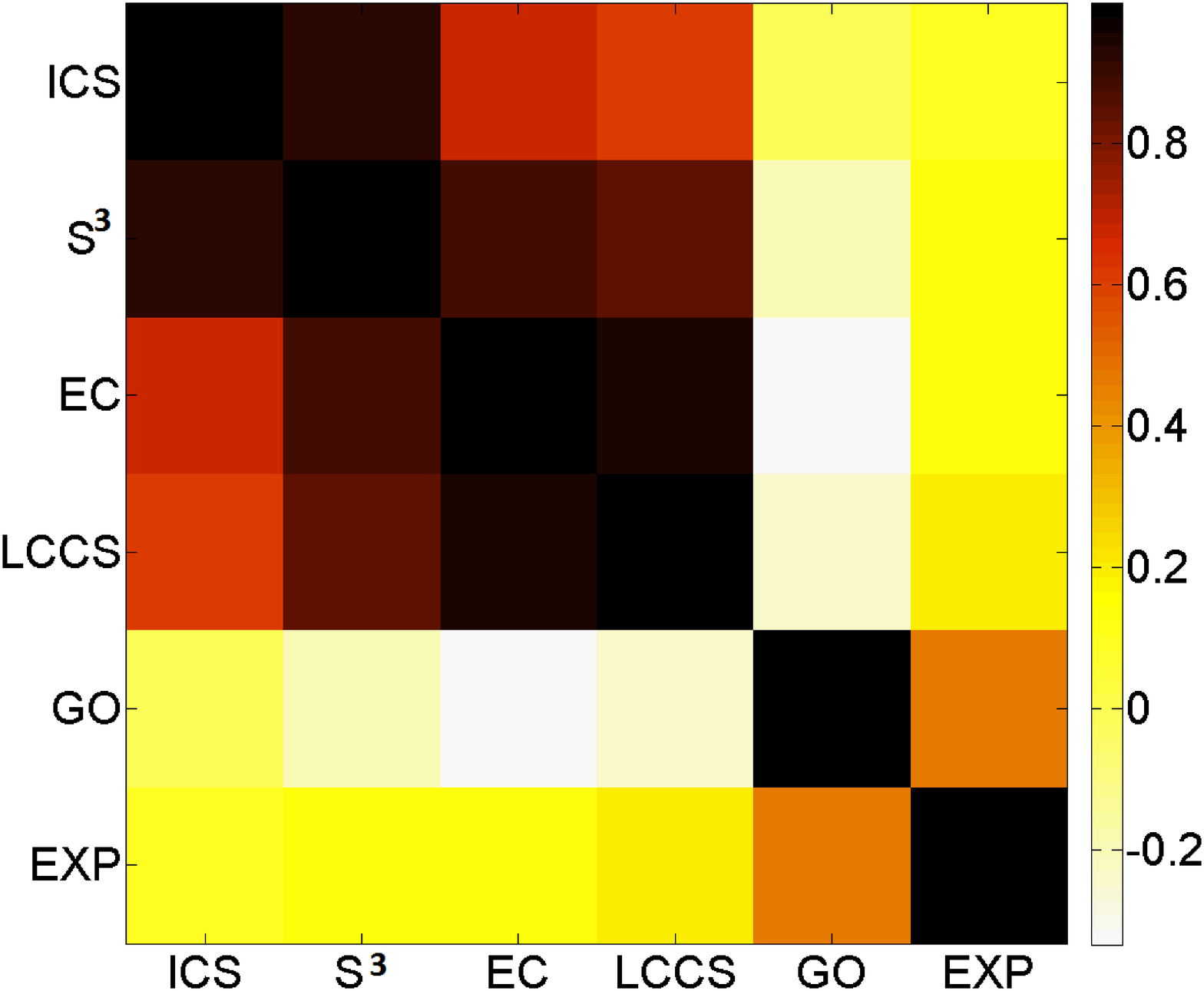}
\caption{Pairwise correlations between different alignment quality
measures for: \textbf{(a)} synthetic networks with known ground
truth node mapping and \textbf{(b)} real networks with unknown node
mapping. Correlations were computed over alignments with the highest
NC scores in panel (a) and over alignments with the highest EC
scores in panel (b) (because we do not known NC scores for
alignments of real networks). Note that color scales for the two
panels are different.}
\label{fig:heat}
\end{figure}

NC significantly correlates with both topological and biological
alignment quality measures (Fig.~\ref{fig:heat} (a)), which is
encouraging. Further, for the synthetic network data set, it is also
encouraging that all other measures significantly correlate well
(Pearson correlation coefficient of at least 0.8), even though we see
some clustering of the topological measures and also of the biological
measures (Fig.~\ref{fig:heat} (a)). Interestingly, each of the two
biological measures, GO and EXP, correlates better with some of the
topological measures (e.g., EC) than with each other.

Unlike for the synthetic network data, for the real network data, the
topological measures now correlate poorly with the biological measures
(Pearson correlation coefficient of at most 0.2; Fig.~\ref{fig:heat}
(b)). Importantly, this implies that for the real network data set it
might be hard to produce an alignment that is of excellent quality
both topologically and biologically. Also, while we again see
clustering of the topological measures, the two biological measures
now correlate weakly (Fig.~\ref{fig:heat} (b)), indicating that the
choice of GO annotation data obtained by experimental evidence code
matters (Section \ref{quality_measures}).

The result differences between the synthetic networks and the real
networks could be due to differences in their properties (Section
\ref{data}).

\section{Conclusions} \label{conclusion}

We have aimed to systematically answer three questions in the context
of MI-GRAAL and GHOST network aligners: 1) what is the contribution of
each method's NCF and AS to the alignment quality, 2) how much
sequence versus topology information should be used within NCF when
generating an alignment, and 3) how large the size of the
neighborhoods of the compared nodes from different networks should be.
Our results represent a set of general recommendations for a fair
evaluation of any global network alignment method, not just MI-GRAAL
and GHOST.

Genomic sequence alignment has revolutionized our biomedical
understanding. Biological network alignment has already had similar
impacts. And given the tremendous amounts of biological network data
that continue to be produced, network alignment will only continue to
gain importance. The hope is that it could lead to new discoveries
about the principles of life, evolution, disease, and therapeutics.

\section*{Acknowledgements}

We thank Dr. R. Patro and Dr. C. Kingsford for their assistance with
running GHOST. This work was supported by the National Science
Foundation CCF-1319469 and EAGER CCF-1243295 grants.

\bibliographystyle{plain}
\bibliography{network_bib}

\end{document}